\documentclass[12pt,a4paper]{article}%
\pdfoutput=1

\usepackage[T1]{fontenc}
\usepackage[utf8]{inputenc}
\usepackage{amsmath}
\usepackage{amssymb}
\usepackage{amsthm}
\usepackage[ 
    ]{hyperref}
\usepackage{mathtools}
\usepackage{graphicx}
\usepackage{booktabs}
\usepackage{caption}
\usepackage{subfig}
\usepackage{slashed}
\usepackage{listings}
\usepackage{slashed}
\usepackage{rotating}
\usepackage{feynmp}
\usepackage{subfig}
\usepackage{geometry}
\usepackage{multicol}
\usepackage{dsfont}
\usepackage[nottoc]{tocbibind}
\usepackage{multicol}

\numberwithin{equation}{section} 
\setcaptionmargin{1cm}
\newcommand{\hhref}[1]{\href{http://arxiv.org/abs/#1}{{\it arXiv:#1}}}

\DeclarePairedDelimiter{\abs}{\lvert}{\rvert}

\begin{document}


\begin{titlepage}
$\quad$
\vskip 4.0cm
\begin{center}
{\huge \bf  Weak Scale Baryogenesis in a Supersymmetric Scenario with R-parity violation} 
\\ 
\vskip 2.0cm {\large
{\bf Fabrizio Rompineve} 
} \\[1cm]
{\it 
 Institut de Th\'eorie des Ph\'enom\`enes Physiques, EPFL, Lausanne, Switzerland 
} \\
{\it and}\\
{\it Institut f\"ur Theoretische Physik, Universit\"at Heidelberg, Philosophenweg 19\\
D-69120 Heidelberg, Germany}\\[5mm]
\vskip 1.0cm
\today
\end{center}
\begin{abstract}

We investigate the connection between R-parity violation (RPV) in supersymmetric models and Baryogenesis. First we discuss in detail the assumptions of a theorem by Nanopoulos and Weinberg on the CP asymmetry generated from the decay of massive particles. In light of this statement, we analyse some interesting models of Baryogenesis through RPV. We then explore, in the context of RPV SUSY, the possibility to generate the baryon asymmetry through the out-of-equilibrium decay of a metastable Weakly Interacting Massive Particle (WIMP), as proposed in \cite{Cui:2012jh}. This setting is also motivated by the observed coincidence between the abundances of dark and baryonic matter. In this framework, we propose two models of out-of-equilibrium decay of a would-be cold relic, and compute the associated CP asymmetry. With a TeV-scale parent, the observed baryon abundance can be reproduced in these models when the stop is in the multi-TeV region. Furthermore, annihilation of the metastable particle into SM states must be characterised by a very weak coupling, $g\sim 10^{-2}$ and by a heavy mediator $m_{med}\sim 10$ TeV. These models can also accommodate stop masses far from the TeV scale, at the expense of weakening the explanation of the $\Omega_{B}-\Omega_{DM}$ coincidence.
\end{abstract}

\end{titlepage}


\section{Introduction}

The long-awaited observation of a Higgs boson with mass of about 125 GeV completed the discovery of the particle content of the Standard Model (SM). However the unexplained separation between the Fermi scale and any higher relevant scale in Nature (e.g. the Planck scale, the GUT scale, etc.) makes the SM unnatural. This \emph{hierarchy problem} might be solved by some new physics, which would be either weakly or strongly coupled. In the former picture, Supersymmetry (SUSY) and its minimal realizations (MSSM, NMSSM) are the main possible extensions of the SM. The most general supersymmetric and renormalizable superpotential which can be  written using the field content of the MSSM contains terms that violate the lepton (L) and the baryon (B) numbers. Those interactions would lead to proton decay and other phenomenological catastrophes. A discrete symmetry, known as R-parity, is therefore introduced in the MSSM in order to forbid those dangerous terms. As a consequence, the lightest supersymmetric particle (LSP) must be stable and therefore provides a good candidate for Dark Matter (DM).

Despite its advantages, it is reasonable to ask whether R-parity is really a solid theoretical assumption, especially since the LHC has not found evidence of superpartners to date. R-parity violation (RPV) (see e.g. \cite{Barbier:2005pr}) could play a role in explaining the absence of these signals (see \cite{Carpenter:2006hs}, \cite{Graham:2012th}), because it would allow superpartners to have only SM particles as decay products.\footnote{For instance, in \cite{Graham:2014vya} the authors show how bounds on superpartner masses can be lowered when RPV is considered.} Furthermore the original motivation for R-parity is weakened by the fact that higher order operators that respect R-parity can induce proton decay, if the scale of SUSY breaking is low (see \cite{Brust:2012jh} for a discussion).\footnote{In this case either L or B must be imposed directly as symmetries of the Lagrangian. Such symmetry can be limited to the SM part of the Lagrangian, such as in models where the Higgs boson is a slepton and an R-symmetry coincides with lepton number (see e.g. \cite{Riva:2012hz}).} RPV is associated to superpotential terms which violate L or B :
\begin{align}
\label{eq:rpv}
 W_{\slashed{L}}&=\frac{1}{2}\lambda^{ijk}L_{i}L_{j}\bar{e}_{k}+\lambda^{'ijk}L_{i}Q_{j}\bar{d}_{k}+\mu^{'i}L_{i}H_{u}\\
 W_{\slashed{B}}&=\frac{1}{2}\lambda^{''ijk}\bar{u}_{i}\bar{d}_{j}\bar{d}_{k}.
\end{align}
If only one among $W_{\slashed{L}}$ and $W_{\slashed{B}}$ is added to the MSSM, then its couplings can escape the bounds coming from proton decay, because the latter involves both $\lambda^{'}_{ijk}$ and $\lambda^{''}_{ijk}$. In any case the RPV couplings involving only the first two generations are usually required to be small by other constraints. It is interesting and important that such a pattern of RPV can be naturally obtained in the framework of some general paradigms: e.g. in Partial Compositness (see \cite{KerenZur:2012} for a discussion on RPV in PC) and Minimal Flavor Violating SUSY (see \cite{Nikolidakis:2007fc, Csaki:2011ge} for the formulation of the MFV SUSY ansatz, and \cite{Krnjaic:2012aj, Csaki:2013we} for models in which MFV is implemented in the RPV sector), where one can also reasonably assume the B-violating couplings to be much larger than the L-violating ones. The possibility to avoid proton decay and other phenomenological constraints even in models with RPV, makes the latter interesting for model building (see \cite{Franceschini:2013ne} for more models of RPV).

A very appealing aspect of RPV is that the $\slashed{B}$ couplings $\lambda^{''}_{ijk}$ may play a role also in the explanation of the Baryon Asymmetry in the Universe (BAU), conventionally defined as:
\begin{equation}
\eta=\frac{n_{B}-n_{\bar{B}}}{n_{\gamma}}
\end{equation}
where $n_{B}$, $n_{\bar{B}}$, $n_{\gamma}$  are the number densities of baryons, antibaryons and photons. Its observed value, $\eta\approx 10^{-10}$ \cite{Komatsu:2008hk}, is usually addressed in the framework of baryogenesis, where it is generated dynamically at some energy scale, rather than being imposed as an initial condition. 

Baryogenesis, as established by Sakharov \cite{Sakharov:1967dj}, must involve B, C and CP violation, as well as out-of-equilibrium dynamics. It is now understood that the last two conditions are probably not sufficiently satisfied in the Standard Model (SM) (see \cite{Strumia} for a discussion, and references therein): this motivates the quest for a mechanism to generate the BAU in frameworks beyond the SM. In this regard, as we have just mentioned, R-parity violating SUSY may offer suitable scenarios. 

Supersymmetric scenarios are known to provide also possible solutions to the Dark Matter (DM) puzzle. For instance, a TeV-scale LSP neutralino in the MSSM realizes the paradigm of a Weakly Interacting Massive Particles (WIMP) as DM. The key feature of a weakly interacting cold relic is that its abundance matches the order of magnitude of the DM abundance (see e.g. \cite{Profumo:2013yn}). In this framework dark matter and baryons have very different characteristic interactions. Nonetheless, the abundances of DM and Baryons are astonishingly similar: $\Omega_{DM}\approx 5\Omega_{B}$.

There are by now several frameworks which address this coincidence. Asymmetric Dark Matter (\cite{Nussinov:1985xr}, see also \cite{Kaplan:2009ag} for recent models), where a (B-L) asymmetry is generated at high energies and transferred to the dark matter sector, predicts DM masses in the range $5-15$ GeV, out of the region which allows for the WIMP miracle. The latter is kept as an important ingredient in recent work: e.g. \cite{McDonald:2011zza}, in which however the ratio of dark matter and baryon abundances depends parametrically on the temperature at which an initial asymmetry is transferred to some B-charged scalar fields; \cite{Cui:2011ab} (see also \cite{Bernal:2012gv}), where the baryon abundance is determined by WIMP DM annihilation but certain parameters have to be adjusted in order to avoid washout; \cite{Davidson:2012fn}, where the baryon asymmetry is generated via leptogenesis at the TeV scale. The latter proposal is also sensitive to washout processes.

In the direction of explaining the  $\Omega_{DM}\approx 5\Omega_{B}$ coincidence, while preserving the paradigm of WIMP DM, it has been recently proposed \cite{Cui:2012jh} that the BAU might be generated by the out-of-equilibrium decay of a metastable WIMP. In this framework dark matter is assumed to be a stable cold relic. In the approximation of infinite lifetime, a TeV-scale metastable particle can reproduce the WIMP miracle. If baryons are generated from its decay, their abundance is related to the would-be abundance of the parent. Setting aside potentially dangerous wash-out effects, there are then two conditions for $\Omega_{B}$ to be naturally close to $\Omega_{DM}$, according to this mechanism. First of all a large CP asymmetry must be obtained in the decay of the metastable WIMP, otherwise baryons and antibaryons are produced in similar amounts and the model does not match the observed baryon asymmetry. Then a certain hierarchy among the masses and couplings of the DM particle and the baryon parent is needed, because $\Omega_{B}$ is suppressed by at least three orders of magnitude, corresponding to the splitting between the TeV scale of the WIMPs and the proton mass. In a concrete realisation of this idea, the required hierarchy measures how plausible this mechanism is. The ideal case is represented by a CP asymmetry of $O(0.1)$, requiring a difference only of $O(1)$ among the masses and couplings of the WIMPs to match the observed BAU and naturally explain the $\Omega_{B}-\Omega_{DM}$ coincidence.

In \cite{Cui:2012jh}, the authors propose a mechanism to obtain such a large CP asymmetry, and hint at the possibility of implementing it in the framework of R-parity violating Split SUSY (\cite{DimoSplit}, \cite{GiudiceSplit}, see \cite{Arvanitaki:2012ps} for its Mini version). As we have mentioned, the connection between the WIMP miracle and the baryon asymmetry is the main idea of this proposal.\footnote{But see also \cite{Cui:2013bta}, where the author studies a concrete model of baryogenesis from a WIMP in Mini-Split SUSY, setting aside the $\Omega_{DM}-\Omega_{B}$ coincidence. This model is reviewed in Sec.~\ref{sub:reviewcui}.}

The aim of this paper is twofold: first of all we would like to analyse the assumptions underlying an important result by Nanopoulos and Weinberg \cite{Nanopoulos:1979}, concerning the $B$ asymmetry produced in the decay of a massive particle. This result can be used to check existing models of Baryogenesis through the R-parity violating couplings.
We would then like to describe two tentative realisations of the model proposed in \cite{Cui:2012jh} in RPV SUSY with heavy squarks. Assuming that baryogenesis is caused by the decay of a TeV-scale would-be cold relic, these implementations reproduce the observed baryon abundance when the lightest sfermion is not far from the TeV scale, in agreement with the usual estimate based on Naturalness. Heavier squarks are allowed, though they require a heavier baryon parent and a less plausible ranges of annihilation parameters.

This paper is organized as follows. In Sec.~\ref{sec:nanweinberg} we review the useful statement of Nanopoulos and Weinberg, providing also a proof which stresses the importance of one of the assumptions of the result.
In Sec.~\ref{sec:rpvreview} we briefly discuss the experimental bounds on the RPV couplings, and review some literature on RPV baryogenesis and leptogenesis, showing that some of the existing models seem not to take into account the result of Nanopoulos and Weinberg. 
In Sec.~\ref{sec:barwimp} we focus on the scenario proposed in \cite{Cui:2012jh}, which is shortly explained in \ref{sub:reviewcui}. In Sec.~\ref{sub:attempts} we describe two possible incarnations of the model of \cite{Cui:2012jh} in the framework of RPV SUSY. 

\section{The Nanopoulos and Weinberg theorem}
\label{sec:nanweinberg}
 
Long ago Nanopoulos and Weinberg have obtained a simple but important result concerning those baryogenesis scenarios in which the BAU arises from the out-of-equilibrium decay of a massive particle \cite{Nanopoulos:1979}. Their statement is the following:

\bigskip
\noindent \emph{Consider the decay of a particle $X$ involving $\slashed{B}$ interactions. Assume that $X$ is stable when the $\slashed{B}$ interactions are switched off. Then at first order in the baryon number violating interactions, the decay rate of $X$ into all final states with a given value $B$ of the baryon number equals the rate for the corresponding decay of the antiparticle $\bar{X}$ into all states with baryon number $-B$, that is: $\Gamma(X\rightarrow f)=\Gamma(\bar{X}\rightarrow \bar{f})=\bar{\Gamma}$, where $f$ denotes the n-particles final state.}
\bigskip

\noindent The result is a consequence of the unitarity of the $S$ matrix and of the CPT theorem. Let us split the $S$ matrix in two parts:
 \begin{equation}
 \label{eq:smatrix}
  S=S_{0}+iT_{\slashed{B}},
 \end{equation}
 where $S_{0}=1+iT_{0}$ does not violate $B$, and $T_{\slashed{B}}=\sum_{n}\lambda^{n}T^{(n)}$ is the $\slashed{B}$ transition matrix, with $\lambda$ being a real dimensionless expansion parameter. The transition amplitude from the state $X$ to the final state $f$, with $B(X)\neq B(f)$, at first order in $\lambda$ and at all orders in the B-preserving interactions, is given by:
 \begin{equation}
  \langle f|S| X\rangle=i\lambda\langle f|T^{(1)}|X\rangle+O(\lambda^{2}).
 \end{equation}
 The S-matrix is unitary, which means $S^{\dagger}S=SS^{\dagger}=\mathds{1}$. Expanding the S-matrix using (\ref{eq:smatrix}), at first order in $\lambda$, we find:
 \begin{equation}
 \label{eq:unitarity}
  T^{(1)}=S_{0}T^{(1)\dagger}S_{0}.
 \end{equation}
 Therefore:
 \begin{equation}
  \langle f|T^{(1)}|X\rangle=\langle f|S_{0}T^{(1)\dagger}S_{0}|X\rangle.
 \end{equation}
 Now, since X is stable under B-preserving interactions by hypothesis, $S_{0}|X\rangle=|X\rangle$, so we have:
 \begin{equation}
 \label{eq:weinberg1}
  \langle f|T^{(1)}|X\rangle=\langle f|S_{0}T^{(1)\dagger}|X\rangle=\sum_{h}\langle f|S_{0}|h\rangle\langle h|T^{(1)\dagger}|X\rangle.
 \end{equation}
 We can now use the CPT theorem: $\langle h|T^{(1)\dagger}|X\rangle=\langle \bar{X}|T^{(1)\dagger}|\bar{h}\rangle$, where the bar denotes the CP conjugate state. Therefore \ref{eq:weinberg1} becomes:
 \begin{equation}
   \langle f|T^{(1)}|X\rangle=\sum_{h}\langle \bar{X}|T^{(1)\dagger}|\bar{h}\rangle\langle \bar{h}|S_{0}|\bar{f}\rangle=\sum_{h}[\langle\bar{h}|T^{(1)}|\bar{X}\rangle]^{\dagger} S_{0,\bar{h}\bar{f}}
 \end{equation}
 The decay rate is obtained squaring the transition amplitude and integrating on the phase space:
 \begin{equation}
 \label{eq:decayrate}
  \Gamma_{X\rightarrow f}=\lambda^{2}\sum_{f}\int d\Phi_{f}\abs{\langle f|T^{(1)}|X\rangle}^{2}=\lambda^{2}\sum_{h,g}[\langle \bar{h}|T^{(1)}|\bar{X}\rangle]^{\dagger}[\langle\bar{g}|T^{(1)}|\bar{X}\rangle]\sum_{f}\int d\Phi_{f}S_{0,\bar{h}\bar{f}}S_{0,\bar{g}\bar{f}}^{\dagger}.
 \end{equation}
 Let us now complete the proof using again the unitarity of the $S$-matrix: $\sum_{f}\int d\Phi_{f}S_{0,hf}S_{0,gf}^{*}=\int d\Phi_{g}\delta_{hg}$. It is clear that this equality is valid only if the baryon number of the intermediate states $g$ equals the one of the final states $f$, because $S_{0}$ is the S-matrix obtained from the B-preserving interactions. Furthermore, by CPT, the mass of a particle is equal to that of its antiparticle, so that $ d\Phi_{f}=d\Phi_{\bar{f}}$. Since we are summing over the states $g$, with $B(g)=B(f)$, we can rename $g=f$ at the end of the calculation, obtaining the result:
 \begin{equation}
 \label{eq:result}
  \Gamma_{X\rightarrow f}=\Gamma_{\bar{X}\rightarrow \bar{f}}.
 \end{equation}
From the definition of the CP asymmetry in the decay of a massive particle: $\epsilon_{CP}\equiv\frac{\Gamma-\bar{\Gamma}}{\Gamma+\bar{\Gamma}}$, with $\bar{\Gamma}\equiv\Gamma_{\bar{\chi}\rightarrow\bar{f}}$, we straightforwardly obtain the following corollary of the Nanopoulos-Weinberg theorem:
\bigskip

\noindent \emph{In order to generate $CP$ asymmetry from the decay of a massive particle $X$, which is stable in the limit in which the $\slashed{B}$ interactions are switched off, one must consider diagrams that are at least second order in the baryon number violating coupling.}
\bigskip

\noindent Let us comment a bit further on the last statement. Suppose that the massive particle $X$ can decay through a B-preserving interaction. With the notation of the discussion above, this means $S_{0}|X\rangle\neq |X\rangle$, and implies $|X_{0}^{in}\rangle\neq|X_{0}^{out}\rangle$. If this is the case, then the result \ref{eq:result} is not valid, i.e. $\bar{\Gamma}\neq\Gamma$. Therefore, using a B-preserving decay channel, we may be able to build diagrams, involving only one power of the $\slashed{B}$ coupling, which indeed provide a CP asymmetry. 

Finally let us notice that decays involving $L$, rather than $B$, violation obviously obey a corresponding result, obtained by replacing baryon number with lepton number in the statement of the theorem. 

\section{Review on Baryogenesis from R-parity violation}
\label{sec:rpvreview}

In this section we will review some models of baryogenesis from R-parity violation, in light of the Nanopoulos-Weinberg theorem.\footnote{The literature is more concisely reviewed in \cite{Barbier:2005pr}, w/o reference to the result by Nanopoulos and Weinberg.}
Before doing so, let us briefly discuss the constraints on the RPV couplings coming from the BAU (see \cite{Barbier:2005pr} for a review). The point is that the $\slashed{R}$ couplings can erase any baryon asymmetry generated before the ElectroWeak Phase Transition (EWPT). Sphalerons are in equilibrium at high energies \cite{Kuzmin:1985}, in particular in the range $T_{EWPT}\sim 100$ GeV $\lesssim T\lesssim 10^{12}$ GeV. Now, sphalerons preserve $(B-L)$ so that any baryon asymmetry generated before the EWPT survives only if it originates from a $(B-L)$ asymmetry: leptogenesis implements precisely this idea (see \cite{Strumia} for a review). However the $\slashed{R}$ couplings violate $(B-L)$. Therefore a $(B-L)$ asymmetry can be preserved only if the $\slashed{R}$ interactions are out-of-equilibrium after the EWPT, i.e. if $\Gamma_{\slashed{R}}<H(T_{C})$. In particular the strictest bounds on $\Gamma$ are obtained from the decay of squarks and sleptons into two fermions or sfermions (see \cite{Barbier:2005pr} and references therein for a computation of the rates):
\begin{align}
\label{eq:ewptbounds}
 &\Gamma_{\lambda}\lesssim 1.4\times 10^{-2}\abs{\lambda}^{2}\frac{\tilde{M}^{2}}{T}\\
 &\Rightarrow \abs{\lambda}\lesssim 10^{-7} \quad \text{for}\quad \tilde{M}\simeq T\sim T_{C}\\
\end{align}
where $\tilde{M}$ is the mass of the decaying sfermion. The same constraints are valid for all the RPV couplings $\lambda,\lambda^{'},\lambda^{''}$. If $\tilde{M}\sim 1$ TeV, the upper bound is increased by a factor of $3$. Since the sphalerons preserve $B-L_{i}$ for each lepton flavor $i$, ans it is sufficient to have only one $B-L_{i}$ asymmetry after the EWPT the bounds in (\ref{eq:ewptbounds}) are valid for every generation. 

Let us remark here that the bounds (\ref{eq:ewptbounds}) are not valid if the BAU is generated at the weak scale, when the sphalerons are not efficient anymore (see e.g. \cite{Davidson:2008bu} for a concise discussion on the regime of efficiency of the sphalerons).  

We are now ready to review some of the proposed models of baryogenesis through R-parity violation. We will divide them into two categories: those that make use of $\slashed{R}$ couplings to generate a lepton asymmetry, then convert it to a baryon asymmetry through sphalerons, and those that generate the BAU directly. Obviously each one of these models has to satisfy Sakharov's conditions. B and L violation come with the $\slashed{R}$ interactions. Except for the first model, the out-of-equilibrium condition is  satisfied due to the expansion of the universe. The most interesting condition for our analysis is the violation of CP in the out-of-equilibrium decays, because it is constrained by the Nanopoulos-Weinberg theorem. There are two ways to evade it: by building decay diagrams that are at least of second order in the $\slashed{B}$ or $\slashed{L}$ couplings and/or by allowing $B$- or $L$-preserving decay channels. Apart from the particle content, the models can therefore be classified by looking at how they evade the statement by Nanopoulos and Weinberg.

\subsection{$\slashed{R}$ leptogenesis}

We will first discuss some models in which the baryon asymmetry is obtained through the interactions obtained from the superpotential term $W_{\slashed{L}}$.

Let us start by the model proposed by Masiero and Riotto \cite{MasieroRiotto}. They study the generation of a lepton asymmetry through $\slashed{R}$ interactions at the EWPT, then convert it to the BAU using sphalerons, which therefore must be still efficient just after the EWPT. This is in contrast with what is usually assumed, i.e. that the sphalerons play a rôle only above the EWPT: however this is possible if one or more singlet superfields are added to the Higgs sector of the MSSM. The aforementioned bounds (\ref{eq:ewptbounds}) are therefore not valid in this case. The lepton asymmetry is generated by the $\slashed{CP}, \slashed{L}$ decay of the lightest neutralino $\tilde{\chi}^{0}$, which is assumed to be the LSP. The EWPT is assumed to be of first order, and in particular proceeds via percolation of subcritical bubbles, originated by thermal fluctuations.
We will not discuss in detail the dynamics of bubble nucleation and collision (see \cite{MasieroRiotto}): the important point is that in the collisions the energy of the bubble can be released through the direct production of particles, whose distribution will be far from equilibrium. Neutralinos are produced through this mechanism, which therefore provides for Sakharov's condition. 
Their decay proceeds through the $\Delta L=1$ channel $\tilde{\chi}\rightarrow t,l_{i},\bar{d}_{k}$, with a virtual stop: this decay involves the complex coupling $\lambda^{'}_{i3k}$. The one-loop diagram interfering with the tree level to generate the $CP$ asymmetry involves three powers of the coupling $\lambda^{'}$: it therefore seems to violate one of the hypotheses of the Nanopoulos-Weinberg theorem.
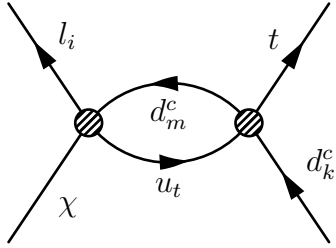
\begin{figure}[]
  \centering
   \begin{fmffile}{effloop}
    \begin{fmfgraph*}(150,90)
     \fmfleft{i1,i2}
     \fmfright{o1,o2}
     \fmf{vanilla,label.side=right,label=$\chi$,tension=2}{i1,v1}
     \fmf{fermion,label.side=right,label=$l_{i}$,tension=2}{v1,i2}
     \fmf{fermion,right=0.5,label.side=left,label=$d_{m}^{c}$}{v2,v1}
     \fmf{fermion,right=0.5,label.side=right,label=$u_{t}$}{v1,v2}
     \fmf{fermion,label.side=left,label=$t$,tension=2}{v2,o2}
     \fmf{fermion,label.side=right,label=$d^{c}_{k}$,tension=2}{o1,v2}
     \fmfblob{10}{v1}
     \fmfblob{10}{v2}
    \end{fmfgraph*}
   \end{fmffile}
   \caption{One-loop diagram interfering with the tree level in the effective theory obtained by integrating out the heavy scalars, in the model proposed in \cite{MasieroRiotto}. Only the first vertex on the left violates the lepton number.}
   \label{fig:effloop}
  \end{figure} 
The authors claim that the $CP$ asymmetry obtained in this model is:
\begin{equation}
\label{eq:mrcp}
 \epsilon_{CP}\approx \frac{1}{16\pi}\frac{\sum_{ikmst}\Im[\lambda^{'*}_{itm}\lambda^{'}_{stm}\lambda^{'*}_{s3k}\lambda^{'}_{i3k}]}{\sum_{ik}\abs{\lambda^{'}_{i3k}}^{2}},
\end{equation}
and, as expected, depends on the phases of the couplings $\lambda_{ijk}^{'}$. The couplings $\lambda^{'}_{i3k}$ turn out to be constrained by two conditions: the process $\tilde{\chi}\rightarrow t,l_{i},\bar{d}_{k}$ must be at equilibrium at the EWPT, when the neutralinos are produced, i.e. $\Gamma_{\tilde{\chi}\rightarrow tl_{i}\bar{d}_{k}}\gtrsim H(T_{C})$ and potentially wash-out processes must be out-of-equilibrium, e.g.  $\Gamma_{l_{i}\bar{d}_{k}\rightarrow t\tilde{\chi}}\lesssim H(T_{C})$. In fact, as the authors argue, the asymmetry is suppressed anyway because some of the neutralinos may thermalize before decaying through the $\tilde{\chi}t\rightarrow\tilde{\chi} t$ scattering.
The lepton asymmetry generated in the decay is finally converted into a baryon asymmetry using sphalerons. The observed value of $\eta$ is obtained for $\lambda^{'}\simeq 8\times10^{-3}$, within the range dictated by the aforementioned bounds, with stop mass $m_{\tilde{t}}\approx 5$ TeV, and $m_{\tilde{\chi}}\approx 500$ GeV.
 
Let us now push the analysis of this model a bit further, using once again the theorem by Nanopoulos and Weinberg.\footnote{I thank M. Nardecchia for a discussion on this analysis of the model.} To this end, let us consider the same decay diagrams generating the lepton asymmetry in the effective theory which is obtained integrating out the heavy squarks and sleptons. The one loop diagram interfering with the tree level is shown in Fig.~\ref{fig:effloop}. It contains only one $L$-violating vertex, because the effective coupling responsible for the scattering process $d^{c}_{m},u_{t}\rightarrow t,d^{c}_{k}$ does not violate $L$. In other words, this diagram is only of first order in the $L$-violating couplings of the effective theory. Furthermore, since $\tilde{\chi}$ is assumed to be the LSP, it is stable under L-preserving interactions. Therefore the hypotheses of the Nanopolous-Weinberg theorem are both respected and the CP asymmetry must vanish. This result is valid in the effective theory. Going back to the full theory, this implies that a CP asymmetry can arise only if suppressed by inverse powers of the mass of the heavy superpartners, in contrast with \ref{eq:mrcp}.

A variation of this model has been studied in \cite{Sarkar1}. Starting from the observation that in many supersymmetric models the mass of the sfermions is of $O(\text{TeV})$, the authors propose to consider the lepton asymmetry generated by the $\slashed{R}$ decays of the sfermions, as they too may be produced in blubble collisions. They then write down all the possible $\slashed{R}$ sfermion decays: at tree level they are mediated by the coupling $\lambda^{'}$ only. At one loop the decays involve the neutralino, so that the diagram has two MSSM couplings and one $\lambda^{'}$. In order to achieve $CP$ asymmetry they have to impose flavor violation in at least one of the MSSM vertices that they are considering, otherwise the imaginary part of the coefficient vanishes by a suitable redefinition of the fields. 
This model differs from the previous one in the way it evades the Nanopoulos-Weinberg Theorem, as the decay diagrams are only first order in the $\slashed{L}$ coupling $\lambda^{'}$. However the sfermions can decay through the MSSM $L$-conserving channels. The good point of this model is that one can obtain a large CP asymmetry using the MSSM couplings. However let us notice a possible tension in this proposal, based again on the theorem by Nanopoulos and Weinberg: in order to produce sfermions in bubble collisions, one is lowering their mass towards the LSP mass; on the other hand, in the limit in which the mass of the sfermions is equal to that of the LSP, the CP asymmetry vanishes, as a consequence of the result by Nanopoulos and Weinberg, because the sfermions are then stable when the $\slashed{B}$ interactions are switched off.

The scenario proposed by Hambye, Ma and Sarkar \cite{HambyeMa} escapes the consequences of the theorem as in the first of the last two models and it involves the bilinear $\slashed{R}$ couplings $\mu_{i}$. The authors consider the decay of the lightest neutralino through the channel $\tilde{\chi}^{0}\rightarrow \tau_{R}^{\mp}h^{\pm}$, where $h$ is a physical Higgs boson. The decay is assumed to proceed only through the $\tilde{B}$ component of one of the neutralino mass eigenstates, which is denote by $W_{3}^{'}$. $\tilde{B}$ couples to $\tau$ and $\tilde{\tau}_{R}$ through the the R-conserving interaction:
\begin{equation}
 \Delta \mathcal{L}_{\tilde{B}\tau\tilde{\tau}}=-\frac{e\sqrt{2}}{cos\theta_{W}}\Big[\bar{\tau}(\frac{1-\gamma^{5}}{2})\tilde{B}\tilde{\tau}_{R}+c.c.\Big],
\end{equation}
and $\tilde{\tau}_{R}$ mixes with $h^{-}=H_{d}^{-}\cos\beta+H_{u}\sin\beta$ because of the $\mu_{\tau}$ coupling and the interactions of the soft SUSY breaking lagrangian of the MSSM. 
The decay $\tilde{\chi}^{0}\rightarrow \tau_{R}^{\mp}h^{\pm}$ is then suppressed by the coefficient of the $\tilde{B}$ component of the mass eigenstate, and can be out-of-equilibrium when the temperature of the universe is below the masses of all superpartners except the neutralinos. Thus the model differs from the previous ones also in the way it satisfies Sakharov's condition.  
CP asymmetry is generated from the interference of the tree level and one-loop diagrams of the decay $\tilde{W}_{3}^{'}\rightarrow \tau_{R}^{\pm}h^{\mp}$, which involves heavier mass eigenstate $\tilde{B}^{'}$:
\begin{equation}
\label{eq:hambyeas}
 \epsilon_{CP}=\frac{\alpha\zeta^{2}}{2c^{2}}\frac{Im[\delta^{2}]}{\abs{\delta}^{2}}\Big[1-\frac{m_{h}^{2}}{M_{\tilde{W}_{3}^{2]}}}\frac{x^{1/2}g(x)}{(1-x)}\Big],
\end{equation}
with $x=\frac{M_{\tilde{W}}}{M_{\tilde{B}}}$, and $g(x)=1+\frac{2(1-x)}{x}[\frac{(1+x)}{x}\ln(1+x)-1]$, $c=\cos\theta_{W}$, $\delta=2\frac{M_{\tilde{H}_{u}}M_{\tilde{H}_{d}}}{\mu}$ and $\zeta$ parametrizes the amount of $\tilde{\tau}_{R}-h^{-}$ mixing.
The one-loop self-energy and vertex correction diagrams are of third order in $\slashed{L}$ couplings, therefore the asymmetry (\ref{eq:hambyeas}) is indeed allowed by the theorem of Nanopoulos and Weinberg. Finally from (\ref{eq:hambyeas}) one can obtain the observed value of $\Delta B$ using the still active sphalerons. An important point of this model is that, in order to obtain a sufficient mixing of $\tilde{\tau}_{R}$ and $h^{-}$, one cannot use just the soft breaking term and the superpotential $\mu_{\tau}$ term: the authors then introduce the non-holomorphic term $H_{d}^{\dagger}H_{u}\tilde{\tau}_{L}^{c}$, which is experimentally unconstrained and can provide for the required mixing. 

\subsection{$\slashed{R}$ baryogenesis}

Let us now review some models in which $\Delta B$ is directly produced through the trilinear $\slashed{R}$ and $\slashed{B}$ coupling $\lambda^{''}_{ijk}$ in (\ref{eq:rpv}). They all generate CP asymmetry by allowing B-preserving decay channels, therefore violating one of the hypotheses of the Theorem of Nanopoulos and Weinberg. The out-of-equilibrium condition is satisfied using the expansion of the universe, i.e.~imposing the condition $\Gamma_{\slashed{B}\text{decay}}<H$. The following models thus differ basically in the fields and the phases used to generate the asymmetry.

Dimopoulos and Hall \cite{DimHall} studied the case in which the CP asymmetry is produced by the out-of-equilibrium decay of the squarks into quarks and antiquarks. They assume the squarks to be produced as decay products of the inflaton field: since their momenta will then be of the order of the inflaton mass $M_{I}$, they will be far from thermal equilibrium at reheating, therefore satisfying Sakharov's condition. They assume $M_{I}>\tilde{m}>T_{R}$, where $\tilde{m}$ is the squark mass. 
B asymmetry is obtained from the decay of the stop $\tilde{t}_{R}$ into quarks, through the $\slashed{R}$ coupling $\lambda^{''}_{322}$. In particular the CP asymmetry is generated from the interference of a two-loop diagram involving the top $t_{R}$ and the gluino $\tilde{g}$ as intermediate states, and the triscalar $a$-term of the MSSM soft term, and the tree level one. Since the two loop diagram involves only one power of the $\slashed{B}$ coupling $\lambda_{322}^{''}$, they need to assume $m_{\tilde{g}}<\tilde{m}$, otherwise the theorem of Nanopoulos and Weinberg would forbid any CP asymmetry from this diagram. The consequences of the theorem are escaped because $\tilde{t}$ is not stable under B-respecting interactions: it decays through the channel $\tilde{t}\rightarrow t,\tilde{g}$. 
The asymmetry is also determined by the complex phase of the $a$-term, which is constrained by the experimental observations on the electric dipole moment of the neutron, $d_{n}$.
Scattering processes such as $(\bar{u}_{i},\bar{d}_{k})\rightarrow (\bar{d}_{j},\tilde{g}(\tilde{\gamma}))$ and/or $(g,\bar{d}_{j})\rightarrow (\bar{u}_{i}\bar{d}_{k})$ can potentially wash out the asymmetry. This effect is avoided if $T_{R}/m_{\tilde{g}}<10^{-2}$, so that one can safely assume $\frac{T_{R}}{M_{I}}<10^{-3}$.
Finally the BAU generated by this mechanism is given by:
\begin{equation}
\label{eq:halleta}
 \frac{\eta}{5\times10^{-10}}\simeq \Big[\frac{R}{1/3}\Big]\Big[\frac{T_{R}/M_{I}}{10^{-3}}\Big]\times\Big[\frac{d_{n}}{2.5\times10^{-25} e cm}\Big]\Big[\frac{\tilde{m}}{300 \text{GeV}}\Big]^{2}\abs{\frac{\lambda^{''}_{322}}{1/3}}^{2}.
\end{equation}
In general, from (\ref{eq:halleta}), one needs $T_{R}<1$ GeV if $M_{I}\simeq 1$ TeV: at fixed $M_{I}$, the smaller $T_{R}$, the larger $d_{n}$, so that one can take $T_{R}$ as low as $O(\text{MeV})$ and have $d_{n}$ close to the experimental bounds.  

Another proposal that exploits the phase coming from the $a$-term of the soft SUSY breaking Lagrangian of the MSSM is the one by Cline and Raby \cite{ClineRaby}. They use the out-of-equilibrium decay of the gravitino to generate the BAU in two steps: first of all they generate a squark-antisquark asymmetry through the $\slashed{CP}$ decay of the gravitino and/or of gauginos coming from gravitino's decay. Then the BAU is obtained through the $\slashed{R}$ and $\slashed{B}$ decays of the squarks and antisquarks. Let us first of all recall that gravitinos are in general considered problematic in cosmology \cite{WeinbergGravitino}. On one hand, if the gravitino is stable then its very weak annihilation rate would cause a relic abundance larger than the critical energy density, unless its mass is $m_{3/2}\lesssim 1$ keV. On the other hand if the gravitino is unstable, then its decay rate goes as $\Gamma_{\tilde{G}}\approx\alpha_{\tilde{G}}\frac{m_{3/2}^{3}}{M_{Pl}^{2}}$, and the decay must occurr early enough not to influence the prediction of Big Bang Nucleosynthesis. Furthermore the entropy release after the decay would wash out any baryon asymmetry. Those problems can be avoided if $m_{3/2}\gtrsim 10$ TeV. However in this case gravitinos decouple from equilibrium very early, at $T\lesssim M_{Pl}$ and decay very late, at $T\sim 1$ MeV, which means that their decay indeed satisfies one of the Sakharov's conditions. Cline and Raby argue that the BAU obtained in the aformentioned two step process is given by:
\begin{equation}
 \eta\simeq \Delta B g_{R}\frac{2a\pi^{2}}{11\zeta(3)}\Big[\frac{T_{R}}{m_{\tilde{G}}}\Big],
\end{equation}
where $a=\pi/30$ and $g_{R}$ is the number of relativistic degrees of freedom at the reheating temperature $T_{R}$. The latter must be high enough such that after inflation the gravitinos dominate the energy density of the Universe, i.e. $T_{R}\gtrsim 10^{15}$ GeV. 
CP asymmetry is generated by the relative phase between the triscalar term $\mathcal{L}_{soft}\supset a \tilde{\bar{t}}\tilde{\bar{b}}\tilde{\bar{s}}$ and the relevant gaugino mass $M_{\lambda}$, and by the interference of the tree level decay of the gauginos (gravitino included) and the one-loop diagram involving the $a$-term with intermediate quarks and squarks. For example, in the case of gluinos:
\begin{equation}
 \frac{\Gamma_{\tilde{g}}-\bar{\Gamma}_{\tilde{g}}}{\Gamma_{\tilde{g}}} \approx \frac{\lambda_{323}^{''}}{16\pi}\frac{\Im(a^{*}M_{\tilde{g}})}{\abs{M_{\tilde{g}}}^{2}}.
\end{equation}
Since only one power of the $\slashed{B}$ coupling $\lambda_{ijk}^{''}$ is used in the loop diagrams, these processes would generate a vanishing CP asymmetry according to the result by Nanopoulos and Weinberg. This conclusion is avoided if the squarks running in the loop are taken to be lighter than the gauginos, as the authors assume, so that the latter are not stable under B-conserving interactions.\footnote{The $\tilde{g}\rightarrow t\tilde{t}^{c}$ channel is also exploited in \cite{Kohri:2009ka} to generate a CP asymmetry when the reheat temperature is very low, $T_{R}\sim 1-10$ MeV. The authors also assume the gauginos to be heavier than the squarks, to violate the hypotheses of the Nanopoulos-Weinberg theorem. In that case however the gauginos are produced by the decay of fields (including the inflaton) which belong to an hidden sector, and the gravitino is considered as a candidate for DM.}

Mollerach and Roulet \cite{Mollerach} have proposed a variant of this model: the baryon asymmetry is still produced by the decay of gluinos, exactly as in the previous model, however the gluinos come from the decay of the superpartners of the Peccei-Quinn pseudoscalar axion, the axino and saxino (the remanining scalar degree of freedom in the axion superfield). In order to violate the hypothesis of the theorem of Nanopoulos and Weinberg, the authors take $m_{saxino}>2m_{\tilde{g}}$, $m_{axino}>m_{\tilde{g}}$ so that the decay channels $s\rightarrow \tilde{g}\tilde{g}$, $\tilde{a}\rightarrow \tilde{g}g$ are allowed. The interfering diagrams are the same as before, but here the superpartners decay at $T\sim 1$ GeV, so that there is no risk for nucleosynthesis, and the required CP asymmetry is smaller than before.  

Let us conclude this brief review with a discussion of the model proposed by Adikhari and Sarkar \cite{Sarkar2}. They consider the out-of-equilibrium decay of a neutralino $\chi^{0}_{1}$, which is taken to be the LSP ($m_{\tilde{\chi}}\sim 100-200$ GeV), into quarks: $\chi^{0}_{1}\rightarrow u_{iR},d_{jR},d_{kR}$. At tree level the process is mediated by a squark, which is taken to be heavier than the neutralino ($m_{\tilde{q}}\sim 250-1000$ GeV), and which decays through the RPV couplings $\lambda_{ijk}^{''}$ into quarks. At one loop there are some box diagrams interfering with the tree level process, involving only one power of the $\slashed{B}$ coupling $\lambda_{ijk}^{''}$. The authors claim that a CP asymmetry is produced by the interference, of order $\epsilon_{CP}\sim \lambda_{ijk}^{''}\lambda_{jik}^{''*}$. However, by the very definition of the LSP, it is clear that $\tilde{\chi}^{0}_{1}$ is stable under the $B$-preserving interaction, and that it becomes unstable only because of the $\slashed{R}$, $\
\slashed{B}$ superpotential term in the second line of (\ref{eq:rpv}). Therefore, by the theorem of Nanopoulos and Weinberg, the box diagrams should give a vanishing contribution to the CP asymmetry, in contrast to the claim of the authors. 

To summarize, in this section we have seen several mechanisms of baryogenesis through RPV: they are active at about the weak scale, therefore they escape the bounds on the RPV couplings that we have discussed at the beginning of this section. However, by discussing two examples in some detail, we showed that some of these proposals seem not to take into account the result by Nanopoulos and Weinberg, concerning the possibility of generating a CP asymmetry at linear order in the $\slashed{B}$ couplings.

\section{Baryogenesis from WIMPs}
\label{sec:barwimp}

Up to now we have reviewed models whose only motivation is to reproduce the BAU. In this section we are going to explore the idea that the mechanism responsible for the BAU might be related to the DM abundance.
 
Let us recall that the relic abundance of cold DM is given by (see e.g. \cite{KolbTurner}, or \cite{Profumo:2013yn} for a recent pedagogical introduction):
\begin{equation}
\label{eq:miracle}
\Omega_{WIMP}\simeq 0.1\frac{\alpha^{2}_{weak}/(TeV)^{2}}{<\sigma_{A}\abs{v}>}\simeq 0.1\Big[\frac{g_{weak}}{g_{WIMP}}\Big]^{4}\Big[\frac{m_{med}^{4}}{m_{WIMP}^{2}\cdot TeV^{2}}\Big],
\end{equation}
where $m_{med}$ is the mass of a heavier mediator. Eq. (\ref{eq:miracle}) is the formal expression of the so-called \emph{WIMP miracle}: the abundance of a weakly interacting species with $m_{WIMP}\sim O$(TeV) matches the order of magnitude of the observed $\Omega_{DM}$.\footnote{However it is clear from (\ref{eq:miracle}) that the observed DM abundance can be obtained also from lighter or heavier particles, with couplings that are smaller or larger than the weak one (see e.g. \cite{Feng:2008ya}). All that is needed to realise the miracle is a cold relic with $\langle\sigma_{A}v\rangle\sim 10^{-2}\text{TeV}^{-2}$.} In the approximation of infinite lifetime, the relic abundance of a metastable WIMP is also given by (\ref{eq:miracle}). As proposed in \cite{Cui:2012jh}, if the BAU is generated by the decay of such a WIMP $\chi$, today's abundance of baryons, in the absence of wash-out effects, is approximately given by:
\begin{equation}
\label{eq:baryonab}
\Omega_{B}\simeq \epsilon_{CP}\frac{m_{p}}{m_{\chi}}\Omega_{\chi}^{\tau\rightarrow\infty},
\end{equation}
where $m_{p}$ is the mass of the proton, and $\epsilon_{CP}\equiv\frac{\Gamma_{\chi\rightarrow f}-\Gamma_{\bar{\chi}\rightarrow \bar{f}}}{\Gamma_{\chi\rightarrow f}+\Gamma_{\bar{\chi}\rightarrow \bar{f}}}$ is the CP asymmetry generated in the out-of-equilibrium decay of $\chi$.
From (\ref{eq:miracle}) and (\ref{eq:baryonab}) one obtains the ratio of the baryon to DM abundances, as a function of the characteristic couplings and masses of the two species of WIMPs:
\begin{equation}
\label{eq:abratio}
\frac{\Omega_{B}}{\Omega_{DM}}\approx \epsilon_{CP}\frac{m_{p}}{m_{\chi}}\Big[\frac{m_{DM}}{m_{\chi}}\Big]^{2}\Big[\frac{g_{DM}}{g_{\chi}}\Big]^{4}.
\end{equation}
According to (\ref{eq:abratio}), the goal of any model inspired by this paradigm is to obtain a large CP asymmetry from the decay of the WIMP, $\epsilon_{CP}\sim O(0.1)$. If that is the case, then the $\Omega_{DM}-\Omega_{B}$ coincidence is explained with $O(1)$ differences in the masses and couplings of a stable WIMP, interpreted as DM, and a metastable one, playing the role of the baryon parent. 
Stability of a certain WIMP species might be due to some symmetry under which the DM particle and the decaying particle are differently charged. We will simply assume that there is a mechanism which enforces stability on a species. It should be noticed that in RPV SUSY the LSP (e.g. the neutralino) is generically not stable, and it is therefore in general not a suitable candidate for DM.\footnote{For instance, in \cite{Arcadi:2013jza} the authors consider a light gravitino as dark matter, and generate baryons and DM from a WIMP, in RPV SUSY.} 

Finally, let us emphasize that there are also other valid candidates to explain the dark matter abundance (e.g. axions). Nevertheless, even ignoring the coincidence of abundances, the study of baryogenesis from WIMPs is \emph{per se} interesting.

\subsection{Review of the general model and of a possible incarnation in Mini-Split SUSY}
\label{sub:reviewcui}

\begin{figure}[t]
\centering
\subfloat[][]{
\includegraphics[scale=0.4]{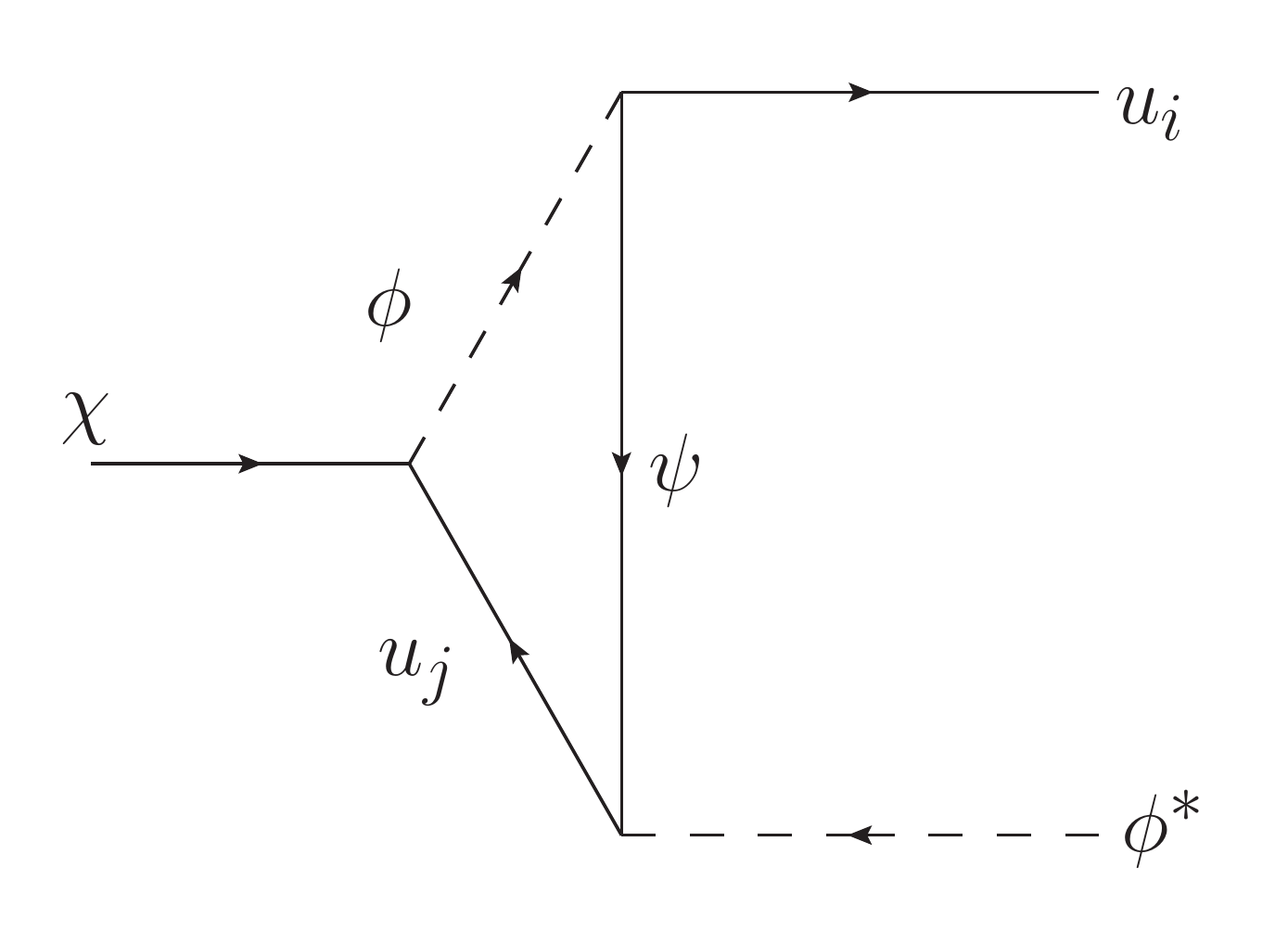}}
\ \hspace{0.5mm} \hspace{0.5mm} \
\subfloat[][]{
\includegraphics[scale=0.4]{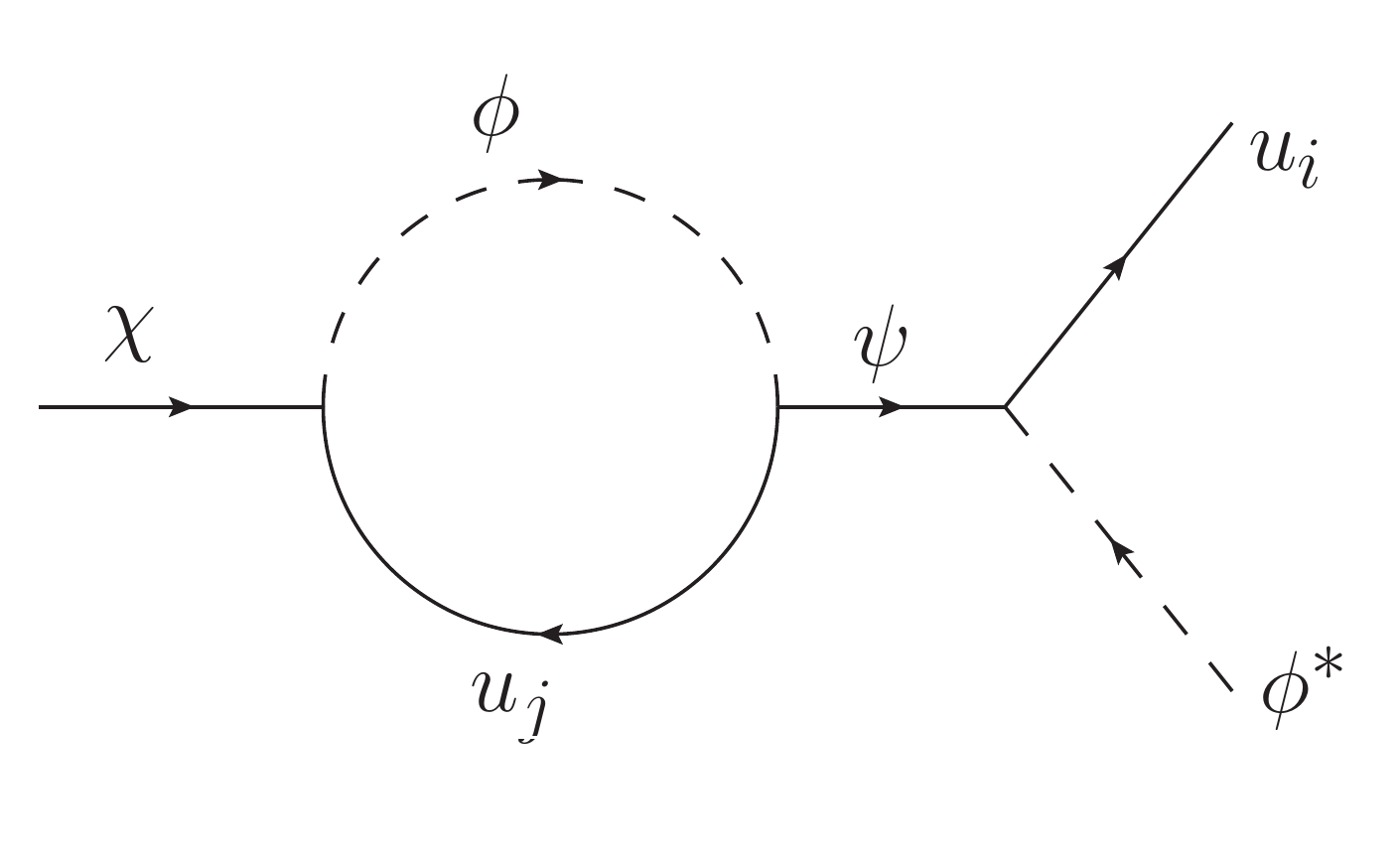}}
\caption{The two loop diagrams that interfere with the tree level decay to generate CP asymmetry. The RPV and $\slashed{B}$ decay of $\phi$ in two quarks is not shown. Figure taken from \cite{Cui:2012jh}.}
\label{fig:cpdiagrams}
\end{figure}

Let us now review a general model which implements the idea that we have just described \cite{Cui:2012jh}. It is described by the Lagrangian:
\begin{equation}
\label{eq:model}
 \mathcal{L}=\mathcal{L}_{SM}+\lambda_{ij}\phi d_{i}d_{j}+\epsilon_{i}\chi\bar{u}_{i}\phi+M_{\chi}^{2}\chi^{2}+y_{i}\psi\bar{u}_{i}\phi+M_{\psi}^{2}\psi^{2}+h.c.,
\end{equation}
where $u$ is the RH SM quark field, i=1,2,3 is the family index, $\phi$ is a di-quark scalar with the same SM gauge charges as $u$, $\chi\equiv\chi_{B}$ and $\psi$ are Majorana fermions representing two generations of metastable WIMPs. For small values of $\epsilon_{i}$ the field $\chi$ is long-lived. The second term in (\ref{eq:model}) mediates the decay $\phi\rightarrow dd$. Together with the out-of-equilibrium decay $\chi\rightarrow\phi^{*}u$, mediated by the third term, they violate B by $\Delta B=1$.
CP asymmetry is generated by the interference of the tree level and one loop amplitudes of the decay $\chi\rightarrow\phi^{*}u$. 
The matrix element of the decay can be factorized into a coupling constant $c$ and an amplitude $\mathcal{A}$:
\begin{equation}
\mathcal{M}=\mathcal{M}_{0}+\mathcal{M}_{1}=c_{0}\mathcal{A}_{0}+c_{1}\mathcal{A}_{1},
\end{equation}
so that, in the case of massless final states, the CP asymmetry is given by the formula (see e.g. \cite{Davidson:2008bu}):
\begin{align}
\label{eq:cpasymmetry}
\epsilon_{CP}\equiv\frac{\Gamma_{\chi\rightarrow\phi^{*}u_{i}}-\Gamma_{\chi\rightarrow\phi\bar{u}_{i}}}{\Gamma_{\chi\rightarrow\phi^{*}u_{i}}+\Gamma_{\chi\rightarrow\phi\bar{u}_{i}}}=\frac{Im[c_{0}c_{1}^{*}]}{\abs{c_{0}}^{2}}\frac{2\int Im[\mathcal{A}_{0}\mathcal{A}_{1}^{*}]d\Phi^{2}}{\int\abs{\mathcal{A}_{0}}^{2}d\Phi^{2}},
\end{align}
where $d\Phi^{2}$ is the two-body phase space of the final state, and the symbol $*$ denotes hermitian conjugation. The decay rate at tree level is: $\Gamma_{tree}=\frac{\sum_{i}\abs{\epsilon_{i}}^{2}M_{\chi}}{8\pi}$.
The one-loop diagrams contributing to the asymmetry are shown in Fig.~\ref{fig:cpdiagrams}. Since the tree level amplitude is real, the imaginary part in (\ref{eq:cpasymmetry}) comes from the one-loop diagrams, and can be computed using Dimensional Regularization, or cutting rules. Assuming the hierarchy $M_{\psi}\gg M_{\chi}$ we find:
\begin{equation}
\label{eq:ascuisundrum}
\epsilon_{CP}\simeq\frac{1}{8\pi}\frac{Im[(\epsilon_{i}y_{i}^{*})^{2}]}{\sum_{i}\abs{\epsilon_{i}}^{2}}\frac{M_{\chi}}{M_{\psi}},
\end{equation}
in agreement with \cite{Cui:2012jh}.
According to (\ref{eq:ascuisundrum}), a large CP asymmetry is obtained if $y_{i}\sim O(1)$. 

Let us emphasize an important point here: the one-loop diagrams of Fig.~\ref{fig:cpdiagrams} develop an imaginary part when the virtual states $\phi, u_{i}$ go on shell. It is therefore clear that the CP asymmetry generated through the decay of the WIMP vanishes if $\phi$ is heavier than $\chi$, as $\phi$ never goes on shell in this case. This result can be obtained also by the statement of Nanopoulos and Weinberg. Indeed in this model the only $B$ violating couplings are the $\lambda_{ij}$s. Those interactions are not shown in Fig.~\ref{fig:cpdiagrams}: they mediate the decay of $\phi^{*}$. Therefore those diagrams are of first order in the $\slashed{B}$ couplings. From the result by Nanopoulos and Weinberg, a vanishing CP asymmetry is expected, unless $\chi$ is not stable when the $\slashed{B}$ interactions are switched off: this happens only if $\phi$ is lighter than $\chi$.

The couplings $\epsilon_{i}$ are constrained by the out-of-equilibrium requirement: $T_{decay}<T_{freeze-out}$. The decay temperature is obtained by the equality: $\Gamma_{decay}=H(T)$, where $H(T)$ is the Hubble rate. The freeze-out temperature is given by (see e.g. \cite{KolbTurner}):
\begin{equation}
\label{eq:freezeout}
 T_{f}\simeq M_{\chi}\Big[\ln(0.038(g_{*}/g_{*S})^{1/2}M_{\chi}M_{Pl}<\sigma_{A}v>\Big]^{-1}\sim \frac{1}{20}M_{\chi}\sim 10^{2} \text{GeV},
\end{equation}
assuming $M_{\chi}\sim O$(TeV). In order to avoid problems with Nucleosynthesis, one also requires: $T_{decay}>T_{BBN}\sim 1$ MeV, obtaining:
\begin{equation}
\label{eq:boundeps}
10^{-13}\lesssim\abs{\epsilon_{i}}\lesssim 10^{-8}.
\end{equation}
Let us now review a recent attempt at an incarnation of this general model in the MSSM with RPV couplings \cite{Cui:2013bta}. There the baryon parent $\chi$ is identified with the bino $\tilde{B}$, while the other Majorana field $\psi$ is identified with a gaugino (a gluino or a wino). A first model of baryogenesis is obtained by considering the out-of-equilibrium decay of the bino in quarks $\tilde{B}\rightarrow d_{i},d_{j},u_{k}$ through the gauge interactions of the MSSM:
\begin{align}
\label{eq:mssmgauge}
\nonumber \mathcal{L}_{gauge}&=\frac{\sqrt{2}}{2}g^{'}(H^{*}_{u}\tilde{H}_{u}\tilde{B}-H_{d}^{*}\tilde{H}_{d}\tilde{B})+\sqrt{2}g^{'}y_{q_{L/R},i}\tilde{q}_{i}^{*L/R,\alpha}q_{i}^{L/R,\alpha}\tilde{B}\\
&+\sqrt{2}g_{w}\tilde{q}_{i}^{*L/R,\alpha}T^{a}q_{i}^{L/R,\alpha}\tilde{W}^{a}+\sqrt{2}g_{s}\tilde{q}_{i}^{*L/R,\alpha}T^{a}q_{i}^{L/R,\alpha}\tilde{g}^{a}+h.c.
\end{align}
At tree level the decay is mediated by a squark, and involves the RPV coupling $\lambda^{''}_{ijk}$, for which the author takes a universal value $\lambda^{''}$. 
Neglecting for simplicity flavor and CP violation in the squark mass matrices, at one loop there is only one diagram interfering with the tree level one: it involves a gluino, and two virtual squarks. Since the one-loop diagram is of first order in the $\slashed{B}$ couplings $\lambda^{''}$, the gluino is required to be lighter than the bino in order to violate the hypothesis of the Nanopoulos and Weinberg theorem. Indeed if $m_{\tilde{g}}<m_{\tilde{B}}$, then the bino has a B-preserving decay channel $\tilde{B}\rightarrow d_{i},\bar{d}_{i}, \tilde{g}$, also involving a virtual squark. In order for this channel not to suppress the asymmetry, the author assumes $\lambda^{''}\gtrsim O(0.1)$. The CP asymmetry is suppressed by the square of the ratio between the mass of the bino and that of the squarks:
\begin{equation}
\label{eq:ascui}
\epsilon_{CP}=\frac{g_{s}^{2}}{15\pi}Im[e^{i\phi}]\Big[\frac{m_{\tilde{B}}}{m_{sfermions}}\Big]^{2},
\end{equation}
where $\phi$ is the phase of the $\tilde{B}$ mass. 

A model of Leptogenesis is also proposed, using the decay channel $\tilde{B}\rightarrow Q_{j}\bar{d}_{k}$, which interferes with a one loop diagram involving the wino and two sleptons. As the gluino in the first model, the wino must be lighter than the bino to allow the B-preserving decay channels $\tilde{B}\rightarrow L_{i},\bar{L}_{i},\tilde{W}$ and $\tilde{B}\rightarrow H,H^{*},\tilde{W}$. The latter however heavily suppresses the asymmetry, unless $\mu\gg m_{sfermions}$. The CP asymmetry is given again by (\ref{eq:ascui}) with $g_{s}$ replaced by $g_{w}$.

The good point of this incarnation is that it uses only the particle content of the MSSM, and that it allows $m_{sfermions}$ to be up to two orders of magnitudes larger than the TeV scale (fitting in the framework of Mini-Split SUSY \cite{Arvanitaki:2012ps}, where it is also viable to have large $\lambda^{''}$ for a generic flavor structure). The \emph{conditio sine qua non} of the model is that $\mu\gg m_{sfermions}$: if this is not satisfied then the B-preserving channel dominates in the leptogenesis mechanism, and the annihilation $\tilde{B},\tilde{B}\rightarrow H,H^{*}$ suppresses the baryon relic abundance also in the baryogenesis scenario.
The hierarchy $\mu\gg m_{sfermions}$ is phenomenologically possible (as recently pointed out in \cite{ArkaniHamed:2012gw}) but requires $B\mu\approx \mu^{2}$ in order to satisfy the condition for ElectroWeak Symmetry Breaking (EWSB) (see e.g. \cite{Martin:1997ns} for a review). The latter condition can be achieved through the Giudice-Masiero mechanism \cite{Giudice:1988yz}.

Let us remark once again that this incarnation focuses on obtaining the observed $\Omega_{B}$ from a metastable WIMP in the framework of Mini-Split SUSY. The explanation of the $\Omega_{DM}-\Omega_{B}$ coincidence, which is the central result of \cite{Cui:2012jh}, is put aside. In the case of WIMP DM, it is not easy to see how to modify the model to explain the aforementioned fact: e.g., the annihilation process $\tilde{B},\tilde{B}\rightarrow H,H^{*}$ can make the bino freeze out as a hot relic, so that its would-be abundance would not be given by (\ref{eq:miracle}), and $\tilde{B}$ would not inherit the WIMP miracle. 

\subsection{Two realisations in SUSY with heavy sfermions}
\label{sub:attempts}

We will now pursue the quest for incarnations of the general model in a different direction, with $\mu$ not far from the TeV scale, and trying to keep the explanation of the $\Omega_{DM}-\Omega_{B}$ coincidence as a motivation. 
As suggested by the fact that the LHC has not found evidence of light superpartners, we will allow the scalars of the supersymmetric theory, except at least one Higgs doublet, to be heavier than the weak scale.   \emph{A priori}, the low energy spectrum of the theory is constituted by the SM fermions, the Higgs doublets $H_{d},H_{u}$, the higgsinos $\tilde{H}_{u},\tilde{H}_{d}$. TeV-scale gauginos can also be considered. We also add two Majorana WIMPs as components of new chiral superfields $\chi, S$. We will assume that they are heavier than the higgsinos, i.e. $\mu < M_{\chi,S}$, and that they are gauge singlets. Furthermore, we require that at least one of them has a fermionic component which is lighter than the squarks. 
\footnote{The case $M_{\chi}>\tilde{M}$ is briefly discussed in \cite{Cui:2012jh}.} We consider the case in which, after SUSY  breaking, there is a certain hierarchy of masses between the two Majorana fermions: let us take $M_{\chi}<M_{S}$. In particular, we will later focus on $M_{\chi}\sim O(\text{TeV})$. We assume that the scalar component of $\chi$ decouples from the low energy spectrum after SUSY breaking.

Let us then consider the following superpotential terms, as suggested in \cite{Cui:2012jh}:
\begin{equation}
\label{eq:cuisuperpotential}
W\supset\lambda_{3ij}TD_{i}D_{j}+\epsilon\chi H_{u}H_{d}+y_{t}QH_{u}T+M_{\chi}\chi^{2}+\mu H_{u} H_{d}+M_{S}S^{2}+\alpha\chi^{2}S+\beta SH_{u}H_{d},
 \end{equation}
 where the superfields $T, D_{i}, D_{j}$ contain the charge conjugated fermionic fields $\bar{d}_{i}=d_{iR}^{\dagger}$, while $\chi$ and $S$ are the new chiral superfields. The superfield $S$ contains a singlet scalar which is responsible for the annihilation of $\chi$ into SM states. Let us denote it by $\tilde{S}$. We did not write linear or cubic terms for $\chi$ and $S$ in the superpotential. In global SUSY, linear terms can be removed by redefining the singlets by constant shifts. A cubic term in $\chi$ would provide another annihilation channel: since we assumed that the scalar $\tilde{\chi}$ is decoupled from the low energy spectrum, this would give a negligible contribution to to the total thermal annihilation cross section. We are not interested in the annihilation of $S$.

\begin{figure}[t]
 \centering
 \begin{fmffile}{effvertex}
  \begin{fmfgraph*}(100,70)
   \fmfleft{i1}
   \fmfright{o1,o2,o3}
   \fmf{fermion,label=$\tilde{H}_{u}$}{i1,v1}
   \fmf{fermion,label.side=left,label=$\bar{b}$}{v1,o3}
   \fmf{fermion,label=$\bar{s}$}{v1,o2}
   \fmf{fermion,label.side=right,label=$t_{L}^{\dagger}$}{v1,o1}
   \fmfblob{10}{v1}
  \end{fmfgraph*}
 \end{fmffile}
 \caption{Effective $\slashed{R}$ vertex.}
 \label{fig:effectiv}
\end{figure}
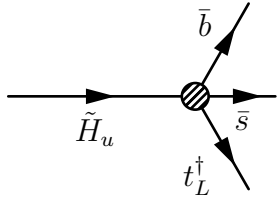

The first term in (\ref{eq:cuisuperpotential}) violates B and R-parity.
From (\ref{eq:cuisuperpotential}) we obtain the following interactions:
 \begin{align}
 \label{eq:int1}
\nonumber  \mathcal{L}_{int}&\supset-\frac{1}{2}[\epsilon H_{u}\tilde{H}_{d}\chi+\epsilon\chi H_{d}\tilde{H}_{u}+c.c.]-\frac{1}{2}[\lambda_{ij}\tilde{\bar{t}}\bar{d}_{i}\bar{d}_{j}+\lambda^{*}_{ij}\tilde{\bar{u}}^{\dagger}\bar{d}_{i}^{\dagger}\bar{d}_{j}^{\dagger}]\\
  &-\frac{1}{2}[y_{t}\tilde{\bar{t}}\tilde{H}_{u}t+y_{t}^{*}\tilde{H}^{*}_{u}\tilde{\bar{u}}^{\dagger}t^{\dagger}]-\frac{1}{2}[y_{t}\tilde{t}\tilde{H}_{u}\bar{t}+y_{t}^{*}\tilde{t}^{*}\tilde{H}^{\dagger}_{u}\bar{t}^{\dagger}]-\frac{1}{2}[y_{t}H_{u}\bar{t}t+y_{t}^{*}H_{u}^{*}\bar{u}^{\dagger}t^{\dagger}].
 \end{align} 
We can now integrate out the stop field, therefore obtaining an effective $\slashed{R}, \slashed{B}$ vertex $\tilde{H}_{u}^{0}\rightarrow t_{L}^{\dagger},\bar{b},\bar{s}$, shown in Fig.~\ref{fig:effectiv}:
\begin{equation}
 \mathcal{A}_{\tilde{H}^{\dagger}\rightarrow b,s,t^{\dagger}}=i\frac{\lambda_{332}^{*}y_{t}}{M_{\tilde{t}}^{2}}[\bar{v}_{\tilde{H}_{u}}P_{L}v_{t}][\bar{u}_{b}P_{R}v_{s}].
\end{equation}
\noindent Only the coupling $\lambda_{332}$ is considered in (\ref{eq:cuisuperpotential}). The remaining independent $\lambda^{''}_{ijk}$ would involve Yukawa couplings $y_{b},y_{s}\ll y_{t}$ and first and second generation squarks. We assume that the latter are not lighter than the stop, so that the only relevant diagrams are the ones with stop mediation. From now on we will write $\tilde{M}\equiv M_{\tilde{t}}$. The coupling $\lambda^{''}_{332}$ could be up to $0(1)$, while $\lambda^{''}_{312},\lambda^{''}_{331}\lesssim 10^{-3}$ (see \cite{Chemtob:2004xr} and references therein). As we mentioned in the introduction, this pattern of third generation dominance is well motivated in certain scenarios.
Let us notice an important feature of the interactions (\ref{eq:int1}): being neutral, $\chi$ mixes with the higgsino $\tilde{H}_{u}$, and decays through the $\slashed{R}, \slashed{B}$ effective coupling in (\ref{fig:effectiv}). However it also decays through the $B$-preserving two-body channel $\chi\rightarrow\tilde{H}_{d}H_{u}$. According to the statement by Nanopoulos and Weinberg, it is therefore possible to obtain a non vanishing CP asymmetry even from loop diagrams involving only one power of the $\slashed{B}$ vertex. 

\begin{figure}[t,b]
 \centering
  \begin{fmffile}{annihilation}
   \begin{fmfgraph*}(150,100)
    \fmfleft{i1,i2}
    \fmfright{o1,o2}
    \fmf{vanilla,label=$\chi$}{i1,v1}
    \fmf{vanilla,label.side=right,label=$\chi$}{i2,v1}
    \fmf{dashes,label=$\tilde{S}$}{v1,v2}
    \fmf{scalar,label.side=left,label=$H$}{v2,o2}
    \fmf{scalar,label=$H^*$}{o1,v2}
   \end{fmfgraph*}
  \end{fmffile}
\caption{Annihilation of $\chi$ into SM states. The same diagram in the gauge eigenstate basis for the Higgs sector is obtained by replacing the labels $H, H^{*}$ with $H_{u}, H_{d}$, and by putting outgoing arrows.}
\label{fig:annihilation}
\end{figure}
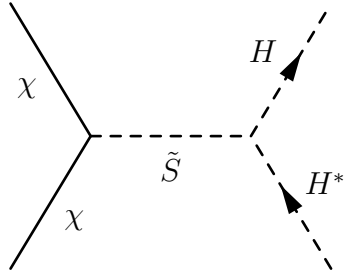

Let us now discuss the would be relic abundance of the metastable fermion $\chi$. Its thermal annihilation into SM states is determined by the last two terms in the superpotential (\ref{eq:cuisuperpotential}). From them one obtains the vertex $\chi, \chi\rightarrow \tilde{S}$, where $\tilde{S}$ is the scalar component of the corresponding superfield, and the triscalar coupling between $\tilde{S}, H_{u}, H_{d}$. Other annihilation channels are subleading: $\chi,\chi\rightarrow H,H^{*}$ through Higgsino mediation involves the very small coupling $\epsilon$, and $\chi,\chi\rightarrow \tilde{H},\tilde{H}^{*}$ has a small cross section due to the heavy mass of the final states. After EWSB the SM Higgs boson arises from the lightest mass eigenstate of the Higgs sector, which we denote by $H$. We assume $m_{H} \ll m_{H^{'}}$. The annihilation process $\chi, \chi\rightarrow H, H^{*}$ is represented diagramatically in Fig.~\ref{fig:annihilation}.
The annihilation cross section is given by:
\begin{equation}
\label{eq:annihilation}
\sigma_{A}(s)=\frac{\abs{\alpha}^{2}\abs{\beta}^{2}}{32\pi}\frac{M_{S}^{2}(s-2M_{\chi}^{2})}{s(s-M_{\tilde{S}}^{2})^{2}}\sqrt{\frac{s-4m_{H}^{2}}{s-4M_{\chi}^{2}}}.
\end{equation}
Its thermal average can be obtained by use of the formula \cite{Gondolo:1990dk} :
\begin{equation}
\label{eq:thaverage}
\langle\sigma_{A}v\rangle=\frac{1}{8M_{\chi}^4TK_{2}^{2}(M_{\chi}/T)}\int_{4M_{\chi}^{2}}^{\infty}ds \sigma_{A}(s)\sqrt{s}(s-4M_{\chi}^{2})K_{1}(\sqrt{s}/T).
\end{equation}
The latter has to be evaluated at the freezeout temperature, which is determined according to the approximation (\ref{eq:freezeout}). We actually approximate (see e.g. \cite{KolbTurner}) (\ref{eq:freezeout}) by replacing $\langle \sigma_{A}v\rangle$ with $(n+1)\sigma_{0}$, where $\sigma_{0}$ is defined by: $\langle \sigma_{A}v\rangle\equiv\sigma_{0}(m/T)^{n}$. In our case $\sigma_{0}\simeq \frac{\abs{\alpha}^{2}\abs{\beta}^{2}}{32\pi}\frac{M_{\chi}^{2}}{M_{\tilde{S}}^{4}}$, with $M_{\chi}<M_{\tilde{S}}$ and $n=1$ for p-wave annihilation. For TeV scale $M_{\chi,S}$ and $\sqrt{\abs{\alpha\beta}}\sim 0(0.1)$, we find $T_{f}\lesssim T_{EWSB}\simeq 246$ GeV. The appropriate value of $g_{*}$ for this temperatures is $g_{*}=75.75$. For $T_{f}\gtrsim T_{EWSB}$, $g_{*}=106.75$. Finally, to a good approximation the would-be relic abundance of $\chi$ is given by:
\begin{equation}
\label{eq:abuchi}
\Omega_{\chi}^{\tau\rightarrow\infty}\simeq\frac{2\cdot10^{9}\text{GeV}^{-1}M_{\chi}}{g_{*}^{1/2}T_{f}M_{pl}\langle\sigma_{A}v(T_{f})\rangle}.
\end{equation}
The latter reproduces (\ref{eq:miracle}) when $T_{f}\simeq M_{\chi}/20$. Let us remark that (\ref{eq:abuchi}) is valid only for a cold relic, i.e. for $M_{\chi}/T_{f}\gtrsim 3$. The latter inequality constrains the parameter space, because the freezeout temperature depends on $M_{\tilde{S}}, \gamma\equiv\sqrt{\abs{\alpha\beta}}$ and $M_{\chi}$. In the spirit of the connection between cold Dark Matter candidates and baryogenesis, we will focus on the case $M_{\chi}/T_{f}\gtrsim 3$, although we will also briefly discuss the changes in the hot relic regime (see the discussion around (\ref{eq:hotrelic})).

Before describing two possible mechanisms of out-of-equilibrium decay of $\chi$, let us make two  quick remarks.
First of all, we do not discuss potentially dangerous wash-out effects, such as inverse decay and baryon violating scatterings. Indeed in this framework the temperature at which the baryon parent decays is always lower than the freezeout temperature. This condition, as explained in \cite{Cui:2012jh}, where an extensive discussion of wash-out processes is presented, should be sufficient to avoid the aforementioned effects. Concerning sphalerons, we have already mentioned that they are effective until $T\sim 10^{2}$ GeV. For a TeV-scale WIMP the freezeout temperature is usually given by $T_{f}\sim \frac{M_{WIMP}}{20}\lesssim 10^{2}$ GeV. In the more general case $T_{f}\gtrsim T_{EWSB}$, the decay temperature $T_{D}$ can always be taken below the region where sphalerons are effective. This can be done by appropriately choosing the coupling $\epsilon$ in the range (\ref{eq:boundeps}) (see also (\ref{eq:bounds2}) in the next subsection).

Secondly, the constraints that are recently discussed in \cite{Barry:2013nva} on $\lambda^{''}_{ijk}$ and squark masses in models of baryogenesis do not affect our discussion. There the authors consider the baryon asymmetry to be generated by some physics which differs from the RPV couplings $\lambda^{''}$, at or above the weak scale. In the framework that we consider, as we have just mentioned, the BAU is introduced at $1 \text{MeV}~\lesssim T_{BAU}\lesssim 10^{2}$ GeV and through the RPV couplings $\lambda^{''}$. Nevertheless the models that we investigate also lead to the generic prediction of displaced vertices at the LHC, as we will comment in Sec.~\ref{sec:conclusions} (see \cite{Graham:2012th} for a recent discussion of displaced vertices from SUSY, \cite{Aad:2011zb} for experimental searches).

\subsubsection{1st realisation}
\label{sub:first}

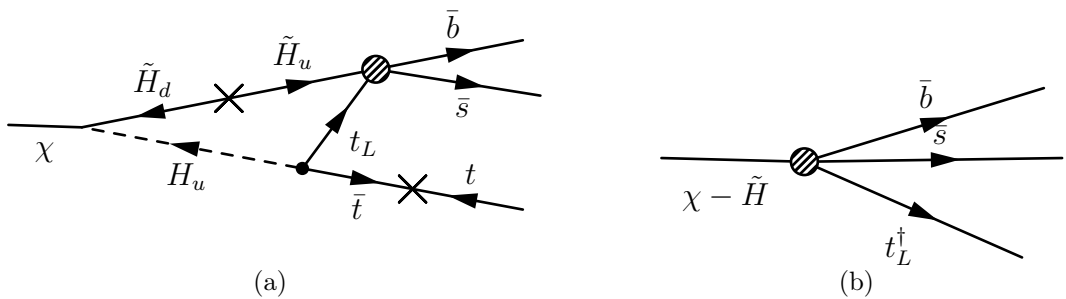
\begin{figure}[b]
 \centering
 \subfloat[][]
 {\begin{fmffile}{cploop}
  \begin{fmfgraph*}(200,100)
    \fmfleft{i1}
    \fmfright{o1,o2,o3,o4,o5,o6}
    \fmf{vanilla,tension=5,label=$\chi$}{i1,v1}
    \fmf{scalar,label.side=left,label=$H_{u}$}{v3,v1}
    \fmf{fermion,label.side=right, label=$\tilde{H}_{d}$}{v4,v1}
    \fmfv{decoration.shape=cross}{v4}
    \fmf{fermion,label.side=left, label=$\tilde{H}_{u}$}{v4,v2}
    \fmf{phantom,tension=2}{v2,t2,o5}
    \fmf{phantom,tension=2}{v3,t,o2}
    \fmfdot{v3}
    \fmffreeze
    \fmf{fermion,label.side=left,label=$\bar{b}$}{v2,o5}
    \fmf{fermion,label=$\bar{s}$}{v2,o4}
    \fmf{fermion,label=$t_{L}$}{v3,v2}
    \fmf{fermion,label=$\bar{t}$}{v3,v6}
    \fmf{fermion,label=$t$}{o2,v6}
    \fmfblob{10}{v2}
    \fmfv{decoration.shape=cross}{v6}
  \end{fmfgraph*}
 \end{fmffile}}
 \ \hspace{2mm} \hspace{4mm} \
 \subfloat[][]
 {\begin{fmffile}{cptree}
   \begin{fmfgraph*}(150,75)
    \fmfleft{i1}
    \fmfright{o1,o2,o2,o3,o4,o5,o6}
    \fmf{vanilla,label=$\chi-\tilde{H}$,tension=5}{i1,v1}
    \fmf{fermion,label.side=left,label=$\bar{b}$}{v1,o5}
    \fmf{fermion,label=$\bar{s}$}{v1,o3}
    \fmf{fermion,label=$t_{L}^{\dagger}$}{v1,o1}
    \fmfblob{10}{v1}
   \end{fmfgraph*}
  \end{fmffile}}
 \caption{One loop and tree level diagram interfering to generate the CP asymmetry. The mass insertions in the loop diagram are used to show the correct direction of the arrows. The virtual state running in the loop is a mix of $\tilde{H_{u}}$ and $\tilde{H_{d}}$.}
 \label{fig:loop}
\end{figure}

Let us first consider the CP asymmetry generated by the interference between the tree level diagram of the decay $\tilde{\chi}-\tilde{H}_{u}\rightarrow t_{L}^{\dagger}, \bar{b}, \bar{s}$, and the one loop one, both shown in Fig.~\ref{fig:loop}.
The one loop diagram contains the $\slashed{B}$ coupling only once, but $\tilde{\chi}$ decays first through the channel $\chi\rightarrow\tilde{H}_{d} H_{u}$, which is $B$-preserving.
In the approximation of massless final states the tree level three-body decay rate is given by:
\begin{align}
\label{eq:treeamp2}
\Gamma_{tree}=\frac{1}{M_{\chi}}\frac{\abs{\epsilon}^{2}\abs{v}^{2}\sin^{2}\beta\abs{y_{t}}^{2}\abs{\lambda_{332}}^{2}}{2^{9}\cdot 3\pi^{3}}\big(\frac{M_{\chi}}{\tilde{M}}\big)^{4}.
\end{align}
According to eq.~(\ref{eq:cpasymmetry}), we then have to compute the imaginary part of the one-loop decay amplitude represented diagramatically in Fig.~\ref{fig:loop}. In principle, there are three possible cuts contributing to it, because the intermediate states can all go on-shell. The two cuts passing through the propagator of the higgsino give a divergent contribution when $x_{i}\rightarrow 0$, with $x_{i}=m_{i}/M_{\chi}$, $i=\mu,h,t$. The other cut, crossing the propagators of the Higgs and of the top quark, gives a vanishing contribution for $x_{i}\rightarrow 0$. Therefore, for $x_{i}\neq 0$, the term associated to it will be subleading in $x_{i}$ compared to those ones coming from the other two cuts. Therefore we neglect it in the computation.

\begin{figure}[t,b]
\centering
\includegraphics[scale=0.6]{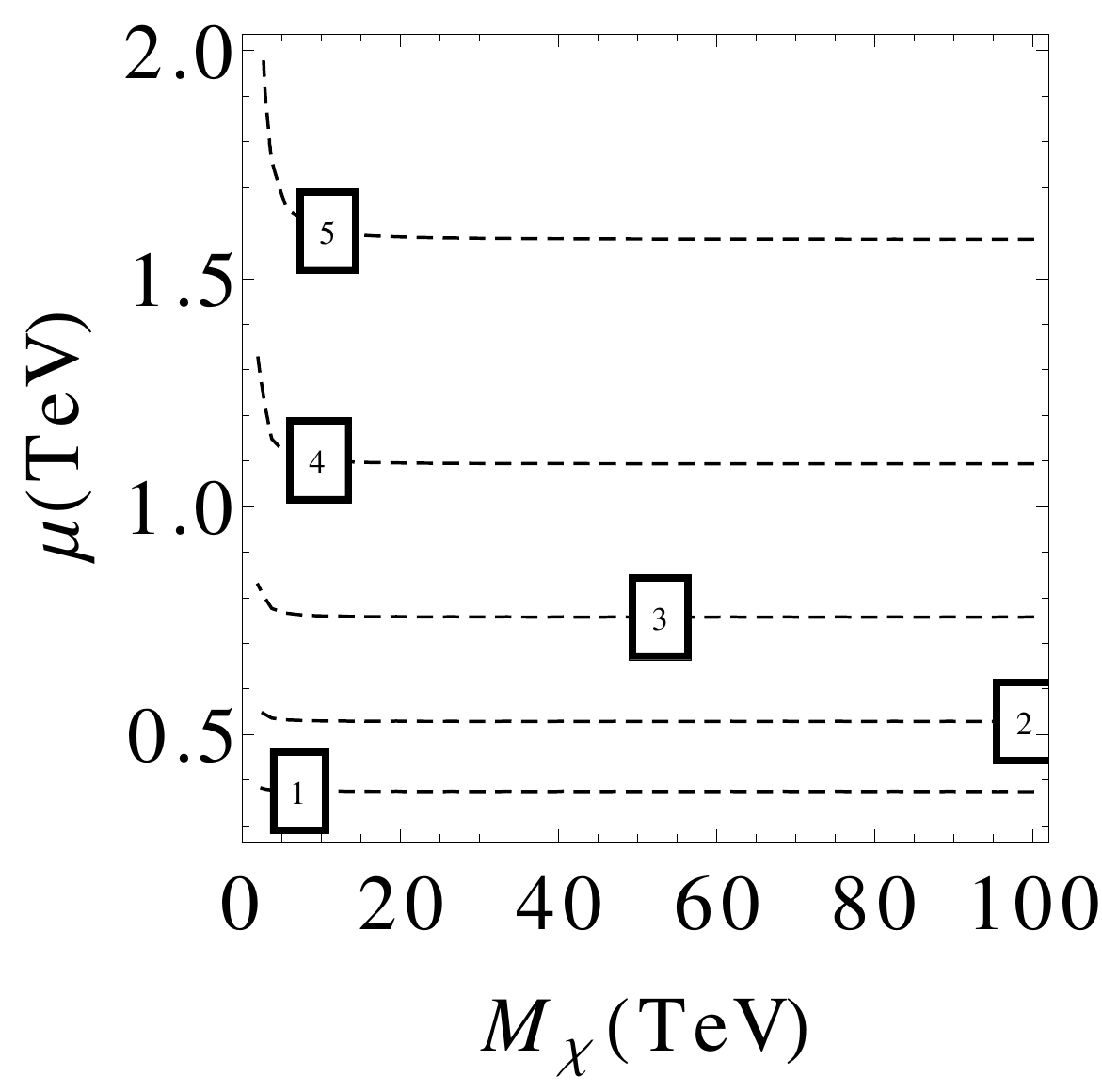}
\caption{Contour plot of $f(x_{\mu},x_{t},x_{h})$ as a function of $\mu$ and $M_{\chi}$.}
\label{fig:contour}
\end{figure}

Using (\ref{eq:cpasymmetry}), we find the following expression for the baryon asymmetry generated by the diagrams in Fig.~\ref{fig:loop}, in the approximation of massless final states, and at second order in $x_{i}$, $i=\mu,h,t$, keeping only the leading terms:
\begin{equation}
\label{eq:cpasy2}
 \epsilon_{CP}=-\frac{Im[c_{0}c_{1}^{*}]}{\sum_{\text{all channels}}\abs{c_{0}}^{2}}\frac{2\int Im[\mathcal{A}_{0}\mathcal{A}_{1}^{*}]d\Phi^{(3)}}{\int\abs{\mathcal{A}_{0}}^{2}d\Phi^{(3)}}\approx\frac{1}{8\pi}\frac{Im\{ \epsilon^{*2}e^{-i\phi_{\mu}}\} y_{t}}{\abs{\epsilon}^{2}\sin\beta}\frac{\abs{\mu} m_{t}}{vM_{\chi}}\frac{f(x_{\mu},x_{t},x_{h})}{A},
\end{equation}
where $\phi_{\mu}$ is the phase of $\mu$, and:
\begin{align}
\label{eq:funct}
f(x_{\mu},x_{t},x_{h})=\Big[-3\frac{x_{h}}{x_{t}}+\frac{1}{3}(2-8\ln\frac{1+\frac{1}{1-2x_{h}^{2}+2x_{\mu}^{2}}}{1-\frac{1}{1-2x_{h}+2x_{\mu}^{2}}}-8\ln x_{t}+12x_{h}^{2}\ln x_{t})\Big],
\end{align}
and $A$ is a suppression factor which, when $\mu\ll M_{\chi}$, is given by:\footnote{In the quantitative analysis performed below we take into account that $\mu\lesssim M_{\chi}$.}
\begin{equation}
\label{eq:suppression}
 A=1+\frac{2^{6}\cdot3\cdot\pi^{2}\tilde{M}^{4}}{\abs{\lambda_{332}^{''}}^{2}\abs{y_{t}}^{2}M_{\chi}^{2}v^{2}\sin^{2}\beta}.
\end{equation}
The cause of this suppression is evident from the 1st line of (\ref{eq:cpasy2}): in the denominator of (\ref{eq:cpasymmetry}) we need to sum the tree level decay rates of all the possible channels. As we have already remarked, apart from the three-body final state, there is also the two-body $\tilde{H}_{d},H_{u}$ channel, which does not involve the mediation of a squark. Therefore its amplitude is enhanced, with respect to that of the decay $\chi-\tilde{H}_{u}\rightarrow t_{L}^{\dagger},\bar{b},\bar{s}$ by a phase space factor, and by a scale factor $\sim\frac{\tilde{M}^{4}}{M_{\chi}^{2}v^{2}}$. A contour plot of the function $f(x_{\mu},x_{t},x_{h})$ is shown in Fig.~\ref{fig:contour} as a function of $\mu$ and $M_{\chi}$, taking $m_{t},m_{h}<\mu$: it is clear that $f\sim$ O(1) in the plotted range. 
The bounds on the coupling $\epsilon$ can be obtained by requiring $T_{BBN}<T_{d}<T_{f}$, as in Sec.~(\ref{sub:reviewcui}). We obtain:
\begin{equation}
\label{eq:bounds2}
10^{-12}\Big[\frac{\tilde{M}}{M_{\chi}}\Big]\lesssim \abs{\epsilon} \lesssim 10^{-7}\Big[\frac{\tilde{M}}{M_{\chi}}\Big],
\end{equation}
similar to (\ref{eq:boundeps}), but with the enhancement factor $\Big[\frac{\tilde{M}}{M_{\chi}}\Big]$. Since $\epsilon$ must be small in order to keep $\chi$ long lived, (\ref{eq:bounds2}) represent a weaker restriction on the splitting between $\tilde{M}$ and the weak scale.
In order to obtain the baryon relic abundance $\Omega_{B}$ from the calculated CP asymmetry (\ref{eq:cpasy2},~\ref{eq:funct},~\ref{eq:suppression}), we combine the latter with (\ref{eq:annihilation},~\ref{eq:thaverage}), (\ref{eq:abuchi}), according to the formula  $\Omega_{B}\simeq \epsilon_{CP}\frac{m_{p}}{m_{\chi}}\Omega_{\chi}^{\tau\rightarrow\infty}$, where $m_{p}\simeq 1$ GeV is the proton mass.

We now present numerical results for the baryon abundance obtained in this model. The analyses performed here and in Sec.~\ref{sub:second} do not represent a complete study of the full parameter space. Rather, they are meant to provide a plausible estimate of allowed regions, assuming typical and/or interesting values for certain parameters. Observationally, there are two main measurements of $\Omega_{B}$: the first one comes from the CMB, $\Omega_{B} h^{2}=0.02207\pm 0.00033$ ($68 \% $, Planck) \cite{Ade:2013zuv}; the second one from BBN $0.021\leq \Omega_{B} h^{2}\leq 0.025$ ($95 \%  $ CL) (\cite{Beringer} and refs. therein). CMB indirect measurements of the Hubble parameter report $H_{0}=100~ h ~\text{km}~\text{s}^{-1}~\text{Mpc}^{-1}=(67.3\pm 1.2)~ \text{km}~ \text{s}^{-1}~\text{Mpc}^{-1}$ ($68 \% $; \textit{Planck}+WP+highL) \cite{Ade:2013zuv}. Astrophysical measurements report $H_{0}=[74.3\pm 1.5(\text{statistical}) \pm 2.1 (\text{systematic})]~ \text{km}~ \text{s}^{-1}~\text{Mpc}^{-1}$ (Carnegie HP) \cite{Freedman}. Therefore the allowed range for the baryon abundance is roughly: $0.035\lesssim \Omega_{B}\lesssim 0.055$.

\subsubsection*{Constraints on the parameter space}

In order to obtain numerical results for the baryon abundance, we take $\lambda^{''}_{332}\simeq O(1)$, $\sin\beta\simeq 1$ and insert the values of $y_{t}, v$ and $m_{t}$.\footnote{As already remarked in Sec.~\ref{sec:rpvreview}, the $\lambda$s do not have to obey the usual bound $\abs{\lambda_{ijk}}<10^{-7}$, as we are studying baryogenesis at or below the weak scale through the RPV couplings. Concerning the choice of $\sin\beta$, we assumed for simplicity $\tan\beta\simeq 10$.} For simplicity, we take $O(1)$ phases of $\epsilon$ and $\mu$. 
At this point there are six parameters left. In order to simplify the analysis, we focus on the case in which the baryon parent has a mass in the TeV region. This choice is well-motivated from the point of view of the cold relic miracle, (\ref{eq:miracle}). It also leads to freezeout temperature not too much above the weak scale, so that it is more easily possible to introduce the baryon asymmetry below the region where sphalerons are effective. We also consider a mild hierarchy with $S$, and take the latter in the multi-TeV region. A larger hierarchy leads to stricter bounds than the ones we will discuss.
As already mentioned, we are forced to take $\mu<M_{\chi}$. We take $\mu$ to be close to $M_{\chi}$ to reduce the suppression of the CP asymmetry due to the B-preserving decay channel $\chi\rightarrow \tilde{H},H$. However we still require the latter to be open, to avoid the Nanopoulos and Weinberg theorem. The baryon abundance then depends on three parameters: the mass of the scalar mediator $M_{\tilde{S}}$, the mass of the stop $\tilde{M}$, and the coupling $\gamma=\sqrt{\abs{\alpha\beta}}$ which determines the cross section of $\chi,\chi\rightarrow H,H^{*}$. 
Since the CP asymmetry is suppressed by the factor $\frac{\tilde{M}^{4}}{M_{\chi}^{2}v^{2}}$, we naively expect that only a small splitting between $\tilde{M}^{2}$ and $M_{\chi}v$ provides the observed $\Omega_{B}$. To be more precise, it is clear that the allowed separation between the scale of the stop mass and the weak scale $v$ depends on the values of $\gamma$ and $M_{\tilde{S}}$. In particular $\Omega_{B}\propto \gamma^{-4}\frac{M_{\tilde{S}}^{4}}{M_{\chi}^{2}\text{TeV}^{2}}$ for $M_{\tilde{S}}\gg M_{\chi}$. 

\begin{figure}[t,b]
\centering
\subfloat[][]{
\includegraphics[scale=0.56]{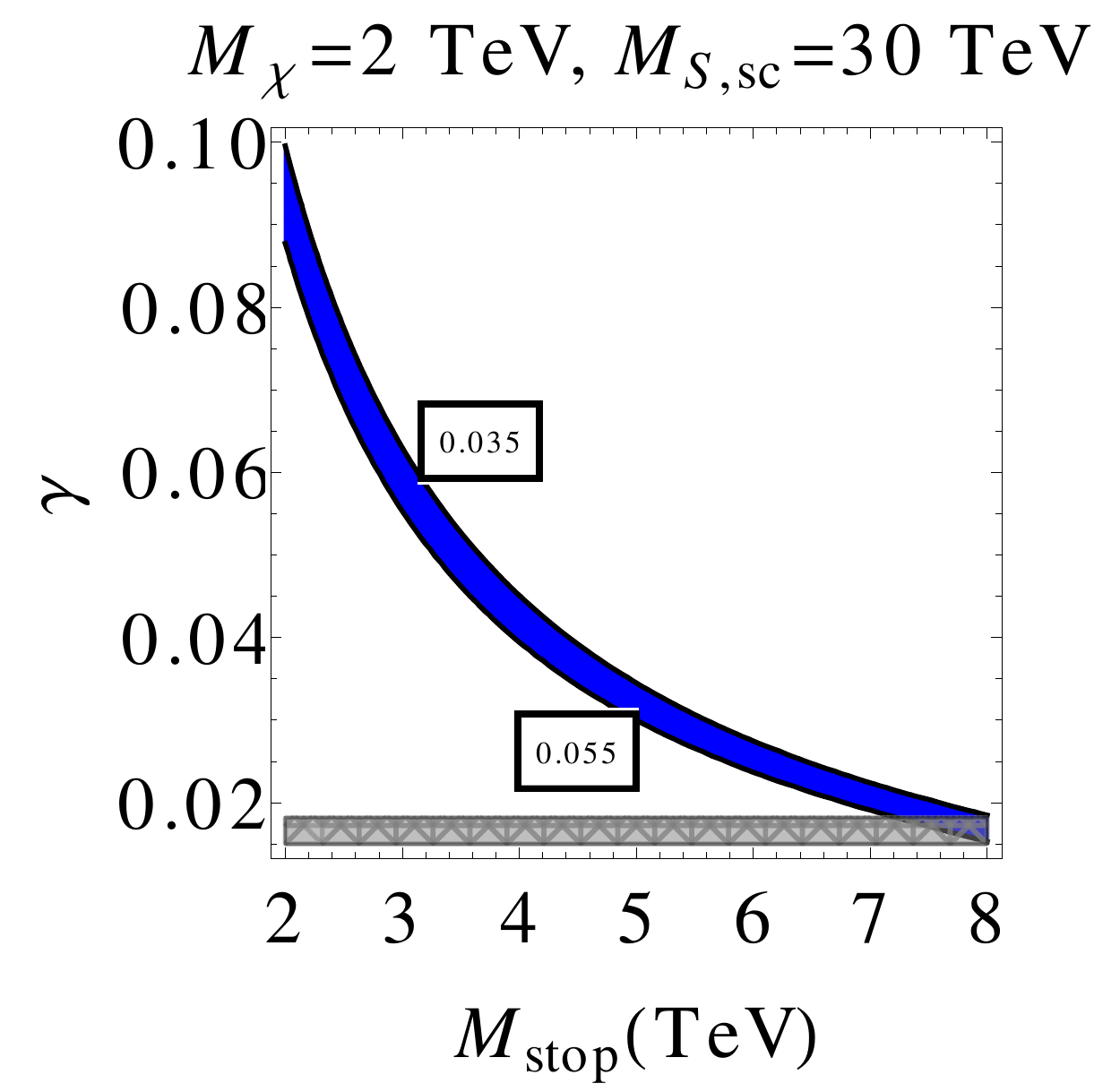}}
\ \hspace{0.5mm} \hspace{0.5mm} \
\subfloat[][]{
\includegraphics[scale=0.55]{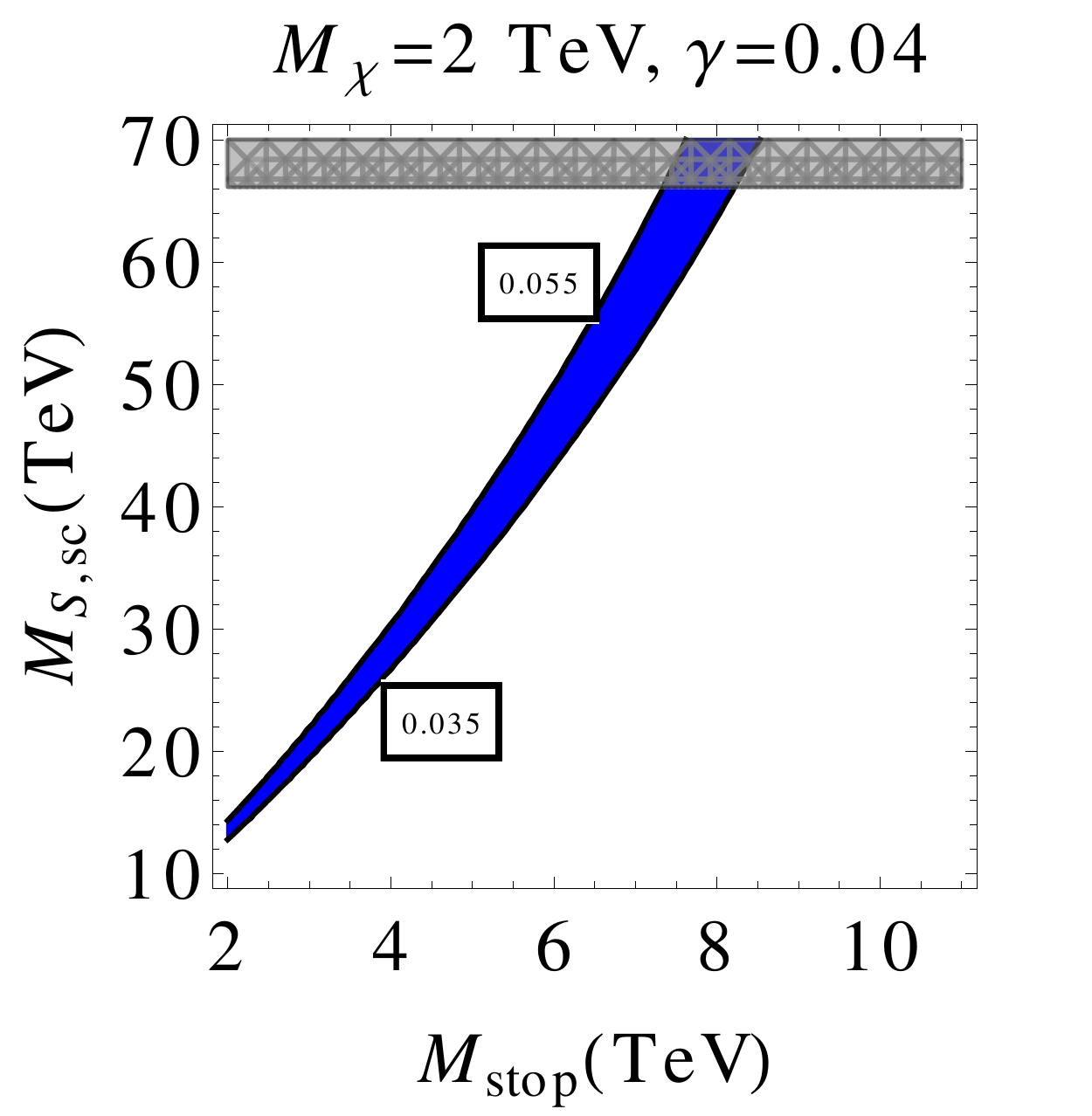}}
\caption{Contour plots of $\Omega_{B}$ as a function of the coupling $\gamma\equiv\sqrt{\abs{\alpha}\abs{\beta}}$, the mediator mass $M_{\tilde{S}}$ and the stop mass $\tilde{M}$. The blue shaded region corresponds to the observed baryon abundance. The gray shaded region is excluded by the condition of cold freezeout: $M_{\chi}/T_{f}\geq 3$. The plots are obtained by taking $M_{\chi}=2$ TeV, $\mu=1.7$ TeV, $M_{S}=4$ TeV and $O(1)$ phases for $\epsilon$ and $\mu$. Furthermore, in a) $M_{S}=30$ TeV is assumed. In b) the coupling constant is fixed at $\gamma=0.04$.}
\label{fig:abundance1}
\end{figure}

In Fig.~\ref{fig:abundance1}, we show the constraints on the parameter space, for $\mu\lesssim M_{\chi}\sim O(\text{TeV})$. We find that, in order to account for the observed baryon asymmetry, the metastable particle $\chi$ has to be generically very weakly coupled, $0.02\lesssim\gamma\lesssim 0.1$, and its annihilation into SM states has to be mediated by a rather heavy scalar, $10~\text{TeV}\lesssim M_{\tilde{S}}\lesssim 60$ TeV. As expected, only a mild hierarchy between the masses of the baryon parent and of the stop is allowed, with $\tilde{M}$ constrained to be in the multi-TeV region. For a fixed value of $M_{\chi}$ it is not possible to arbitrarily tune $M_{\tilde{S}}$ and $\gamma$, because of the requirement that $\chi$ is a cold relic in the limit of infinite lifetime. However the allowed stop mass can be raised by raising $M_{\chi}$ and keeping $\mu$ close to the latter. In particular we found that for $M_{\chi}$ up to $10$ TeV, a stop mass $\tilde{M}\lesssim 20$ TeV is allowed for $\gamma\simeq 10^{-2}$ and $M_{\tilde{S}}\simeq 50$ TeV. It is possible to consider an even heavier baryon parent, such that the upper bound on the stop mass is also raised. However, in this case a smaller coupling and a heavier mediator are required, in particular $\gamma\lesssim 10^{-3}$. Therefore, although the model allows a heavier stop, the required tuning makes the setting less plausible, because the connection with WIMP DM is lost, and so is the explanation of the $\Omega_{B}-\Omega_{DM}$ coincidence (for a discussion on the ranges of annihilation couplings and masses for cold relic dark matter, see e.g. \cite{Profumo:2013yn} and references therein). 

In line with this reasoning, we focus on the constrained parameter space shown in Fig.~\ref{fig:abundance1}. We would like to qualitatively discuss the implications of these constraints on the original motivation of this framework of Baryogenesis. Since we did not assume a specific model of Dark Matter, we consider a generic cross section for the annihilation of a massive particle into SM states:
\begin{align}
\label{eq:dmannihilation}
\nonumber \langle\sigma_{A} v\rangle_{M_{med}>M_{DM}} &\sim g^{4}\frac{M_{DM}^{2}}{M_{med}^{4}}\\
\langle\sigma_{A} v\rangle_{M_{med}<M_{DM}} &\sim \frac{g^{4}}{M_{DM}^{2}},
\end{align}
According to the WIMP miracle (\ref{eq:miracle}), the cross section (\ref{eq:dmannihilation}) must satisfy $\langle\sigma_{A} v\rangle\sim 10^{-2}\text{TeV}^{-2}$. For instance, a particle with $m_{DM}\lesssim 5$ TeV which annihilates through a lighter mediator has the required cross section when $g\sim O(0.5)$. This represents roughly speaking a difference of one order of magnitude for the coupling $\gamma\equiv\sqrt{\abs{\alpha\beta}}$ that is obtained for the baryon parent. However, as we already mentioned, a cold relic can explain the DM abundance with non-WIMP couplings and masses \cite{Feng:2008ya}. Furthermore, the estimate above can be affected by important exceptions to the standard computation of the relic abundance of a cold relic \cite{Griest:1990kh}, e.g. resonances. The argument given here therefore serves only as a naïve way to measure how natural the coincidence between $\Omega_{B}$ and $\Omega_{DM}$ is, according to this model.

\begin{figure}[t,b]
\centering
\includegraphics[scale=0.5]{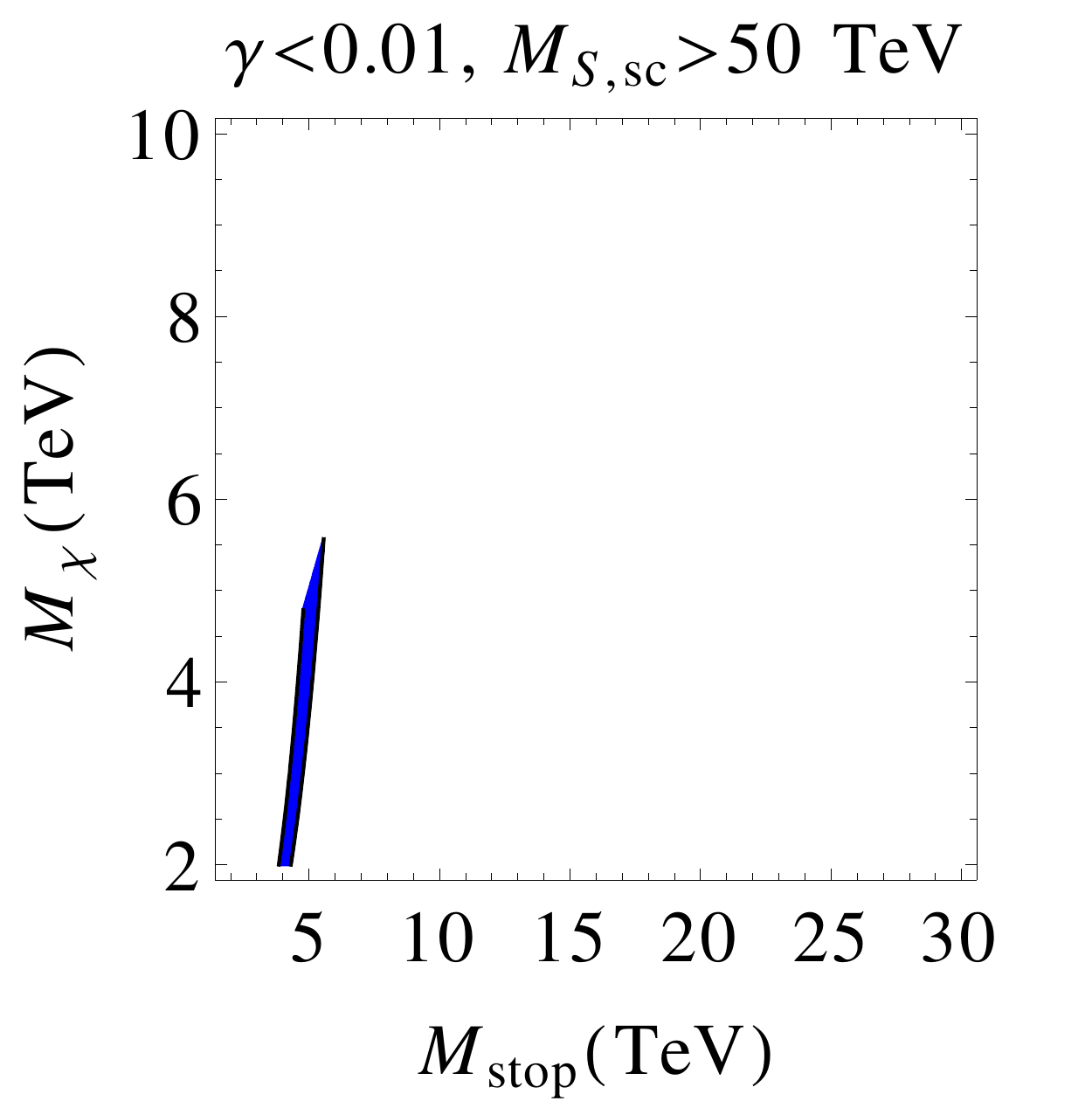}
\caption{Constraints on the parameter space for a would-be hot baryon parent. The blue shaded region corresponds to the observed baryon abundance $0.035\lesssim\Omega_{B}\lesssim 0.055$. The plot has been obtained with $\mu=1.5$ TeV, $\sin\beta=1$. The condition $M_{\chi}<\tilde{M}$ has been imposed as well.}
\label{fig:hotrelic}
\end{figure}

Finally, let us describe how the bounds on the parameter space change when the freezeout temperature is in  the hot relic regime, i.e. $M_{\chi}/T_{f}\lesssim 3$. In this case, the would-be relic abundance of the baryon parent is approximately independent of the details of freezeout (see e.g. \cite{KolbTurner}), and it is given by:
\begin{equation}
\label{eq:hotrelic}
\Omega_{hot}^{\tau\rightarrow\infty}\simeq 1.56\cdot 10^{-1}[\frac{3/4}{g_{*}}]\frac{M_{\chi}}{\text{eV}}.
\end{equation}
Inserting $g_{*}=106.75, M_{\chi}\sim O(\text{TeV})$, one obtains $\Omega_{hot}^{\tau\rightarrow\infty}\simeq 10^{9}$. In Fig.~\ref{fig:hotrelic} we show the allowed region of parameter space in this case. There we fixed $\mu\sim O(\text{TeV})$. Again, we find that a stop mass in the multi-TeV region is allowed, when the baryon parent is at the multi-TeV scale as well.
However we remark that these new bounds are obtained by tuning the parameters $\gamma\lesssim 10^{-2}$ and $M_{\tilde{S}}\gtrsim 50~\text{TeV}$ which determine the annihilation cross section of the baryon parent. Under these conditions, the connection with a cold WIMP Dark Matter is lost, and the explanation of the $\Omega_{B}-\Omega_{DM}$ coincidence is put aside. Even in this less well-motivated case, we find that a multi-TeV-scale stop is required to obtain sufficient baryogenesis, with a TeV-scale baryon parent. Let us also remark that for $M_{\chi}/T_{f}<1$ processes such as inverse decay become important and may wash-out any baryon asymmetry. 

\subsubsection{2nd realisation}
\label{sub:second}

We now illustrate a second possible incarnation of the model of \cite{Cui:2012jh}. In the previous subsection we have built a mechanism which, despite involving only one power of the $\slashed{B}$ coupling, escapes the statement of Nanopoulos and Weinberg because at one loop the WIMP $\chi$ decays through a B-preserving channel. However this is also a problem of that model, as the two-body decay $\chi\rightarrow \tilde{H}_{d},H_{u}$ is much faster than the three-body one $\chi-\tilde{H}_{u}\rightarrow t^{\dagger}_{L},\bar{b},\bar{s}$, and suppresses the CP asymmetry.
This problem can be avoided if we actually use only the B-preserving channel at one loop to generate $\epsilon_{CP}$, mimicking indeed the general model. 

Let us then consider again the superpotential (\ref{eq:cuisuperpotential}). We will now use also the fermionic component of the superfield $S$. We then generate the BAU in two steps: first of all we produce a CP asymmetry in $\tilde{H}_{u}, \bar{\tilde{H}}_{u}$ through the out-of-equilibrium and B-preserving decay $\chi\rightarrow H_{d},\tilde{H}_{u}$. In this model the higgsino $\tilde{H}_{u}$ is the would-be LSP, therefore it decays only through the effective $\slashed{R}$ coupling $\lambda_{332}$. The latter is responsible for converting the CP asymmetry in a baryon asymmetry.

\begin{figure}[b]
 \subfloat[][]{
  \begin{fmffile}{itworksvertex}
   \begin{fmfgraph*}(200,100)
    \fmfleft{i1}
    \fmfright{o1,o2}
    \fmf{vanilla,label=$\chi$,tension=3}{i1,v1}
    \fmf{fermion,label.side=right,label=$\tilde{H}_{u}(p_{1})$}{v2,v1}
    \fmf{scalar,label.side=left,label=$H_{d}(p_{2})$}{v3,v1}
    \fmf{vanilla,label.side=right,label=$S(q)$}{v3,v2}
    \fmf{scalar,label=$H_{d}$}{v2,o2}
    \fmf{fermion,label=$\tilde{H}_{u}$}{v3,o1}
   \end{fmfgraph*}
   \end{fmffile}}
  \ \hspace{2mm} \hspace{8mm} \
  \subfloat[][]{
  \begin{fmffile}{itworksself}
   \begin{fmfgraph*}(200,100)
    \fmfleft{i1}
    \fmfright{o1,o2}
    \fmf{vanilla,label=$\chi$,tension=3}{i1,v1}
    \fmf{fermion,label.side=right,label=$\tilde{H}_{u}(p_{1})$,right=1}{v2,v1}
    \fmf{scalar,label.side=left,label=$H_{d}(p_{2})$,left=1}{v2,v1}
    \fmf{vanilla,label=$S(q)$,tension=3}{v2,v3}
    \fmf{fermion,label.side=right,label=$\tilde{H}_{u}(k)$}{v3,o2}
    \fmf{scalar,label.side=left,label=$H_{d}$}{v3,o1}
   \end{fmfgraph*}
  \end{fmffile}}
  \caption{Vertex and self energy one loop diagrams interfering with the tree level diagram $\tilde{\chi}\rightarrow \tilde{H}_{u}H_{d}$ leading to CP asymmetry.}
  \label{fig:extendedloop}
 \end{figure}
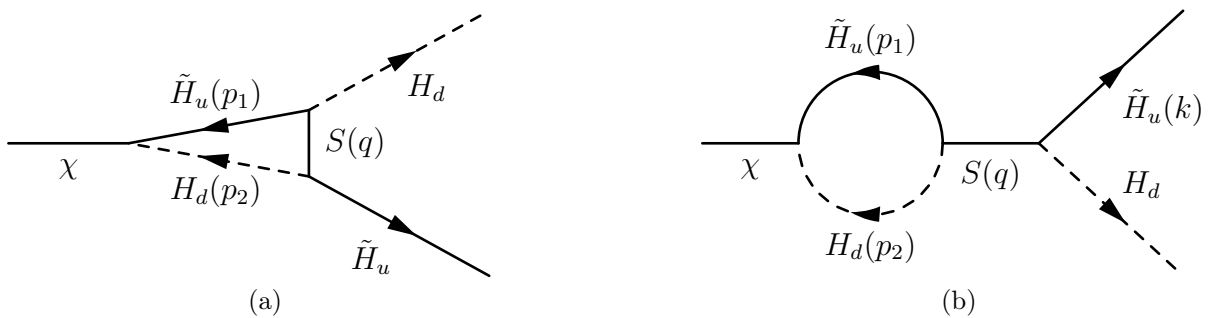

Using the Majorana field $S$ we can build two one-loop diagrams, shown in Fig.~\ref{fig:extendedloop} whose interference with the tree level decay $\chi\rightarrow \tilde{H}_{u}, H_{d}$ generates the CP asymmetry. These diagrams are the analogues of the ones in Fig.~\ref{fig:cpdiagrams}.
The CP asymmetry generated by these diagrams has already been calculated in Section \ref{sub:reviewcui}. Indeed the helicity structure of the diagrams is exactly the same as the one of the diagrams of the minimal model in \cite{Cui:2012jh}. For example, the amplitude represented by the first one-loop diagram is:
\begin{equation}
\mathcal{A}_{vertex}=i\int\frac{d^{4}p_{1}}{(2\pi)^{4}}\int\frac{d^{4}p_{2}}{(2\pi)^{4}}\frac{Tr[P_{R}(\slashed{p}+M_{\chi})P_{R}(-\slashed{p}_{1}+\mu)P_{L}(\slashed{q}+M_{S})P_{L}\slashed{k}]}{(q^{2}-M_{S}^{2})(p_{1}^{2}-\mu^{2})(p_{2}^{2}-m_{h}^{2})}.
\end{equation}
For $M_{S}\gg M_{\chi}$ and in the approximation in which the Higgs and Higgsinos are massless, the CP asymmetry is:
\begin{equation}
\label{eq:cpas2att}
 \epsilon_{CP}\approx \frac{1}{8\pi}\frac{Im\{(\epsilon^{*}\beta)^{2}\}}{\abs{\epsilon}^{2}}\frac{M_{\chi}}{M_{S}},
\end{equation}
Such an asymmetry is large if $\beta\sim O(1)$. The bounds on the coupling $\epsilon$ are the same ones that we discussed at the end of Sec.~\ref{sub:reviewcui}. The CP asymmetry now depends on the relative phase between $\epsilon$ and $\beta$. Let us also notice that a diagram similar to the first in Fig.~\ref{fig:loop} can be obtained using a virtual bino instead of $S$. The bino couples to $H_{u}, \tilde{H}_{u}$ and $H_{d},\tilde{H}_{d}$. We could therefore try to write a more minimal model without considering the field $S$: however in this case we would have the gauge coupling $g^{'}$ instead of $\beta$ and the asymmetry would be further suppressed by a factor of $\sim 10^{-2}$. 

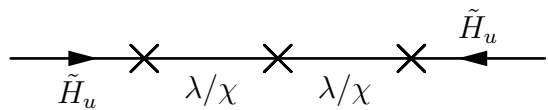
\begin{figure}[b]
 \centering
  \begin{fmffile}{mixing}
   \begin{fmfgraph*}(200,100)
    \fmfleft{i1}
    \fmfright{o1}
    \fmf{fermion,label=$\tilde{H}_{u}$}{i1,v1}
    \fmfv{decoration.shape=cross}{v1}
    \fmf{vanilla,label=$\lambda/\chi$}{v1,v2}
    \fmfv{decoration.shape=cross}{v2}
    \fmf{vanilla,label=$\lambda/\chi$}{v2,v3}
    \fmf{fermion,label=$\tilde{H}_{u}$}{o1,v3}
    \fmfv{decoration.shape=cross}{v3}
    \end{fmfgraph*}
   \end{fmffile}
   \caption{$\tilde{H}^{0}_{u}-\bar{\tilde{H}}_{u}^{0}$ oscillation: the mass insertions in the gaugino propagator denotes the presence of a Majorana mass, while the others the mixing factor.}
   \label{fig:mixing}
   \end{figure}

Despite its simplicity, there is a relevant difficulty associated to this mechanism. Being neutral the higgsinos mix with the neutral gauginos, $\tilde{B},\tilde{W}^{0}$, which are usually both represented by Majorana fields, and with the WIMPs $\chi,S$. This implies that CP conjugate states, $\tilde{H}_{u}$ and $\bar{\tilde{H}}_{u}$, oscillate into one other. The interactions responsible of the mixing are described by the Lagrangian:
\begin{align}
\label{eq:lagrangian}
\nonumber \mathcal{L}_{int}&=-\frac{1}{2}\mu\tilde{H}_{u}\tilde{H}_{d}-\frac{1}{2}M_{\chi}\chi^{2}-\frac{1}{2}M_{S}S^{2}-\frac{1}{2}M_{\tilde{B}}\tilde{B}\tilde{B}-\frac{1}{2}M_{\tilde{W}^{0}}\tilde{W}^{0}\tilde{W}^{0}-\epsilon\chi\tilde{H}_{u}H_{d}
-\beta S\tilde{H}_{u}H_{d}\\
&+\frac{1}{\sqrt{2}}g^{'}H_{u}^{0}\tilde{H}_{u}^{0}\tilde{B}-\frac{1}{\sqrt{2}}gH_{u}^{0}\tilde{H}_{u}^{0}\tilde{W}^{0}.
\end{align}
The amplitude of the oscillation is diagramatically represented in Fig.~\ref{fig:mixing}. 
Qualitatively, we expect that the CP asymmetry in $\tilde{H}_{u},\bar{\tilde{H}}_{u}$ will be washed out by the oscillation if its associated rate $\Gamma_{\tilde{H}_{u}\rightarrow\bar{\tilde{H}}_{u}}$ is larger than the decay rate: $\Gamma_{\tilde{H}_{u}\rightarrow t_{L}^{\dagger},\bar{b},\bar{s}}\approx\frac{\abs{y_{t}}^{2}\abs{\lambda^{''}_{332}}^{2}}{2^{10}\cdot3\pi^{3}}\Big[\frac{\mu}{\tilde{M}}\Big]^{4}\mu$. This semi-quantitative analysis leads again to a restriction on the splitting between the masses of the squarks and $\mu$.

To be more precise, we study the time evolution of the CP asymmetry in (\ref{eq:cpas2att}) with a formalism analogous to the one used to describe CP violation in the decay and mixing of neutral mesons (see \cite{Nir:2005js} for a review). The initial condition on $\epsilon_{CP}$ is given by (\ref{eq:cpas2att}):
\begin{equation}
\epsilon_{CP}^{\chi}=\frac{1}{8\pi}\frac{Im\{(\epsilon^{*}\beta)^{2}\}}{\abs{\epsilon}^{2}}\frac{M_{\chi}}{M_{S}},
\end{equation} 
and there is no further CP violation in the decay of $\tilde{H}_{u}$ ($\bar{\tilde{H}}_{u}$).
The states of an initially pure $|\tilde{H}_{u}\rangle$, or $|\bar{\tilde{H}}_{u}\rangle$ after an elapsed proper time $t$ are denoted  by $|\tilde{H}_{u,phys}(t)\rangle$, $|\bar{\tilde{H}}_{u,phys}(t)\rangle$. The time evolution of these states is described by a $2\times2$ effective Hamiltonian, which is not Hermitian because of the decay $\tilde{H}_{u}\rightarrow t_{L}^{\dagger},\bar{b},\bar{s}$. In the basis $(\tilde{H},\bar{\tilde{H}})$:
\begin{equation}
\label{eq:effhamiltonian}
\mathcal{H}=M-\frac{i}{2}\Gamma=
\begin{pmatrix}
-\frac{i}{2}\Gamma_{\tilde{H}\rightarrow t^{\dagger}_{L},\bar{b},\bar{s}} & m_{M}\\
m_{M} & -\frac{i}{2}\Gamma_{\tilde{H}\rightarrow t^{\dagger}_{L},\bar{b},\bar{s}}
\end{pmatrix}
\end{equation}
The element $\mathcal{H}_{12}$ is a Majorana mass acquired by $\tilde{H}$ because of the mixing with the neutral gauginos $\tilde{B},\tilde{W}^{0}$ and the WIMPs $\chi,S$ described by the diagram in Fig.~\ref{fig:mixing}:
\begin{equation}
\label{eq:maj}
m_{M}=\Big[\frac{g^{'2}v_{u}^{2}}{2M_{\tilde{B}}}+\frac{g^{2}v_{u}^{2}}{2M_{\tilde{W}^{0}}}+\frac{\abs{\epsilon}^{2}v_{d}^{2}}{2M_{\chi}}+\frac{\abs{\beta}^{2}v_{d}^{2}}{2M_{S}}\Big].
\end{equation}
Let us make a brief remark: in principle the effective hamiltonian (\ref{eq:effhamiltonian}) should be a $4\times 4$ matrix, including also the fields $\tilde{H}_{d}$ and its CP conjugate. Then the Dirac mass $\mu$ of (\ref{eq:lagrangian}) would appear, e.g., in the element $\mathcal{H}_{\tilde{H}_{u},\tilde{H}_{d}}$. However notice that the oscillation in Fig.~\ref{fig:mixing} cannot have $\tilde{H}_{d}$ as intermediate state if the VEV of the scalar field $\tilde{\chi}$ vanishes. We will assume $\rangle\tilde{\chi}\langle=0$. Notice also that we did not consider any coupling of $\tilde{H}_{d}$ to quarks in (\ref{eq:cuisuperpotential}). The latter could be relevant for the decay of $\tilde{H}_{u}$, because of the $\mu$ term. However it would involve powers of the Yukawa couplings $y_{d}$ which are much smaller than $y_{t}$. We can therefore study the oscillation of $\tilde{H}_{u}$ and its CP conjugate field without taking $\tilde{H}_{d}$ into account and neglecting subleading contributions. From now on we will also neglect the subindex $u$ in the Higgs superfield.
 
The eigenvalues of the Hamiltonian (\ref{eq:effhamiltonian}) are:
\begin{equation}
\label{eq:eigen}
H_{1,2}=\pm m_{M}-\frac{i}{2}\Gamma.
\end{equation}
and we can write down the time evolution of $|\tilde{H}_{phys}(t)\rangle$ and $|\bar{\tilde{H}}_{phys}(t)\rangle$ as:
\begin{align}
\label{eq:expansion}
&|\tilde{H}_{phys}(t)\rangle=g_{+}(t)|\tilde{H}\rangle-g_{-}(t)|\bar{\tilde{H}}\rangle,\\
&|\bar{\tilde{H}}_{phys}(t)\rangle=g_{+}(t)|\bar{\tilde{H}}\rangle+g_{-}(t)|\tilde{H}\rangle,
\end{align}
where:
\begin{equation}
\label{eq:g}
g_{\pm}\equiv\frac{1}{2}(e^{-iH_{1}t}\pm e^{-iH_{2}t}).
\end{equation}


At the generic time $t$ the system is in a statistical mixture described by a density operator, which in the basis \{$|\psi_{1}\rangle\equiv|\tilde{H}_{phys}(t)\rangle, |\psi_{2}\rangle\equiv|\bar{\tilde{H}}_{phys}(t)\rangle$\} is given by:
\begin{equation}
\label{eq:densitymatrix}
\rho=\begin{pmatrix}
w_{1} & 0\\
0 & w_{2}
\end{pmatrix},
\end{equation}
where $w_{i}$ is the probability to find the system in the state $|\psi_{i}\rangle$. Clearly $w_{1}=\frac{1+\epsilon_{CP}^{\chi}}{2}$, $w_{2}=\frac{1-\epsilon_{CP}^{\chi}}{2}$ and $Tr[\rho]=1$. Using the expansion (\ref{eq:expansion}), we can write the density matrix (\ref{eq:densitymatrix}) in the basis \{$|\phi_{1}\rangle\equiv|\tilde{H}\rangle, |\phi_{2}\rangle\equiv|\bar{\tilde{H}}\rangle$\}:
\begin{equation}
\rho=\begin{pmatrix}
\abs{g_{+}(t)}^{2}(\frac{1+\epsilon_{CP}^{\chi}}{2})+\abs{g_{-}(t)}^{2}(\frac{1-\epsilon_{CP}^{\chi}}{2}) & g_{-}(t)g_{+}(t)^{*}(\frac{1-\epsilon_{CP}^{\chi}}{2})-g_{-}^{*}(t)g_{+}(t)(\frac{1+\epsilon_{CP}^{\chi}}{2})\\
g_{+}(t)g_{-}(t)^{*}(\frac{1-\epsilon_{CP}^{\chi}}{2})-g_{+}^{*}(t)g_{-}(t)(\frac{1+\epsilon_{CP}^{\chi}}{2}) & 
\abs{g_{-}(t)}^{2}(\frac{1+\epsilon_{CP}^{\chi}}{2})+\abs{g_{+}(t)}^{2}(\frac{1-\epsilon_{CP}^{\chi}}{2}).
\end{pmatrix}
\end{equation}
Notice that now $Tr{\rho}=e^{-\Gamma t}$, as the states in (\ref{eq:expansion}) are not normalized.
 
The CP conjugate fields $\tilde{H}, \bar{\tilde{H}}$ do not have final states in common, so the baryon asymmetry at time $T$ is defined by:
\begin{equation}
\label{eq:bas}
\epsilon_{B}(T)=\frac{\int_{0}^{T}dt[Pr_{\tilde{H}_{u}\rightarrow t_{L}^{\dagger},\bar{b},\bar{s}}-Pr_{\bar{\tilde{H}}_{u}\rightarrow t_{L},b_{R},s_{R}}]}{\int_{0}^{T}dt[Pr_{\tilde{H}_{u}\rightarrow t_{L}^{\dagger},\bar{b},\bar{s}}+Pr_{\bar{\tilde{H}}_{u}\rightarrow t_{L},b_{R},s_{R}}]},
\end{equation}
where $Pr_{f}$ denotes the probability of some final state. Now, by definition of the density operator we have:
\begin{align}
Pr_{\tilde{H}_{u}\rightarrow t_{L}^{\dagger},\bar{b},\bar{s}}&=Tr[\rho P_{1}]\\
Pr_{\bar{\tilde{H}}_{u}\rightarrow t_{L},b_{R},s_{R}}&=Tr[\rho P_{2}],
\end{align} 
where $P_{i}$ is the projector on the basis state $|\phi_{i}\rangle$. Using the definitions (\ref{eq:g}) we find:
\begin{equation}
Pr_{\tilde{H}_{u}\rightarrow t_{L}^{\dagger},\bar{b},\bar{s}}-Pr_{\bar{\tilde{H}}_{u}\rightarrow t_{L},b_{R},s_{R}}=Tr[\rho P_{1}]-Tr[\rho P_{2}]=\epsilon_{CP}^{\chi}e^{-\Gamma t}\cos(2m_{M}t),
\end{equation}
while the denominator of (\ref{eq:bas}) gives the normalization factor $\mathcal{N}=\Gamma$.
In the limit $T\rightarrow\infty$ we obtain, from (\ref{eq:bas}):
\begin{equation}
\label{eq:asy2}
\epsilon_{B}^{T\rightarrow\infty}=\frac{\epsilon_{CP}^{\chi}}{1+4\frac{m_{M}^{2}}{\Gamma^{2}}}\simeq\frac{\epsilon_{CP}^{\chi}}{1+\Big[2^{10}\cdot 3\pi^{3}g^{'2}\frac{\tilde{M}^{4}v_{u}^{2}}{M_{\tilde{B}}\mu^{5}}\Big]^{2}},
\end{equation}
where we assume that the dominant contribution to $m_{M}$ comes from the mixing with the bino. 
As expected from the qualitative analysis, for $m_{M}\gg\Gamma$, $\epsilon_{B}\rightarrow\epsilon_{CP}^{\chi}$, i.e. there is no wash-out. The opposite case of total wash-out is obtained for $m_{M}\ll\Gamma$.

Let us now discuss the parameter space in this model. Constraints come from the requirement that $\Omega_{B}\simeq \epsilon_{CP}\frac{m_{p}}{M_{\chi}}\Omega_{\chi}^{\tau\rightarrow\infty}$ matches the observed value of the baryon abundance. Once again, we would like to comment that the analysis that we will provide should be considered as an example of plausible regions of parameter space, rather than as strict bounds.

\subsubsection*{Constraints on the parameter space}

The CP asymmetry obtained in this model, (\ref{eq:asy2}), is suppressed by the ratio $\Big[\frac{\tilde{M}}{{\mu}}\Big]^{8}\Big[\frac{v_{u}^{2}}{M_{\tilde{B}}\mu}\Big]^{2}$, with $v_{u}=v\sin\beta$. Therefore we expect that, in order to preserve the CP asymmetry (\ref{eq:cpas2att}), we need $\tilde{M}\simeq\mu$. However it is not possible to arbitrarily increase $\mu$, because the Higgsinos have to be lighter than the decaying metastable particle in order to violate the hypothesis of the Theorem of Nanopoulos and Weinberg. Assuming $M_{\chi}\sim O(\text{TeV})$ and $\sin\beta\simeq 1$, the observed $\Omega_{B}$ requires $\tilde{M}\sim O(\text{TeV})$. This is the same constraint that characterizes the first model. Furthermore, the suppression (\ref{eq:asy2}) is not ameliorated by taking a heavy bino $M_{\tilde{B}}$: in the case $M_{\tilde{B}}\gg M_{S}$ the dominant contribution to the Majorana mass comes from the last term 
in \ref{eq:maj}, because $\beta\sim O(1)$ and $M_{S}\gtrsim M_{\chi}$ in order to have a CP large asymmetry, (\ref{eq:cpas2att}).
With this guidelines, we take $\mu\lesssim M_{\chi}\sim O(\text{TeV})$, $M_{\tilde{B}}\sim O(3 \text{TeV})$. We also consider the phases of $\epsilon$ and $\beta$ to be $\sim O(1)$, and $M_{S}\simeq 1.5~M_{\chi}$.

\begin{figure}[t]
\centering
\subfloat[][]{
\includegraphics[scale=0.6]{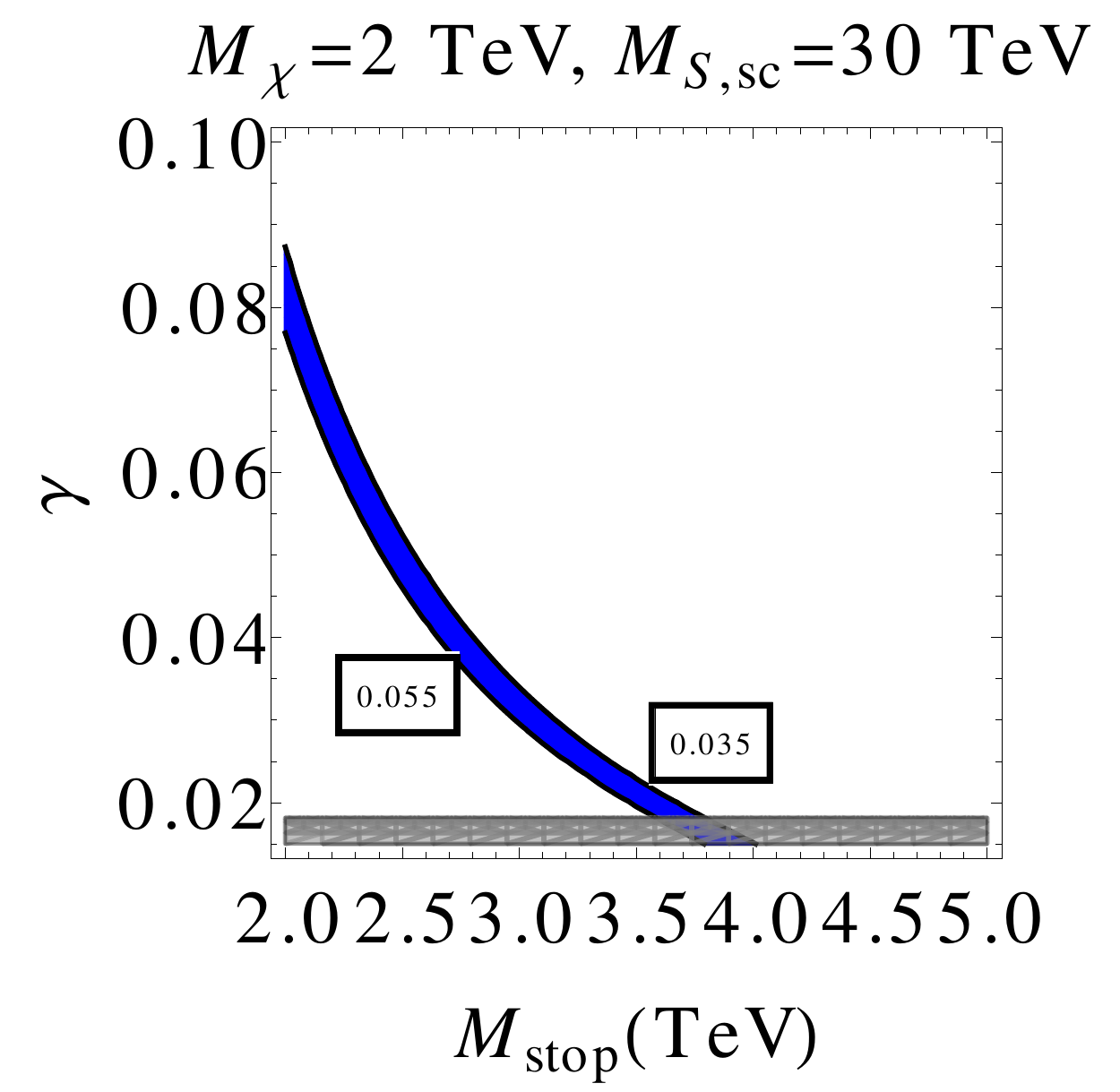}}
\ \hspace{0.5mm} \hspace{0.5mm} \
\subfloat[][]{
\includegraphics[scale=0.57]{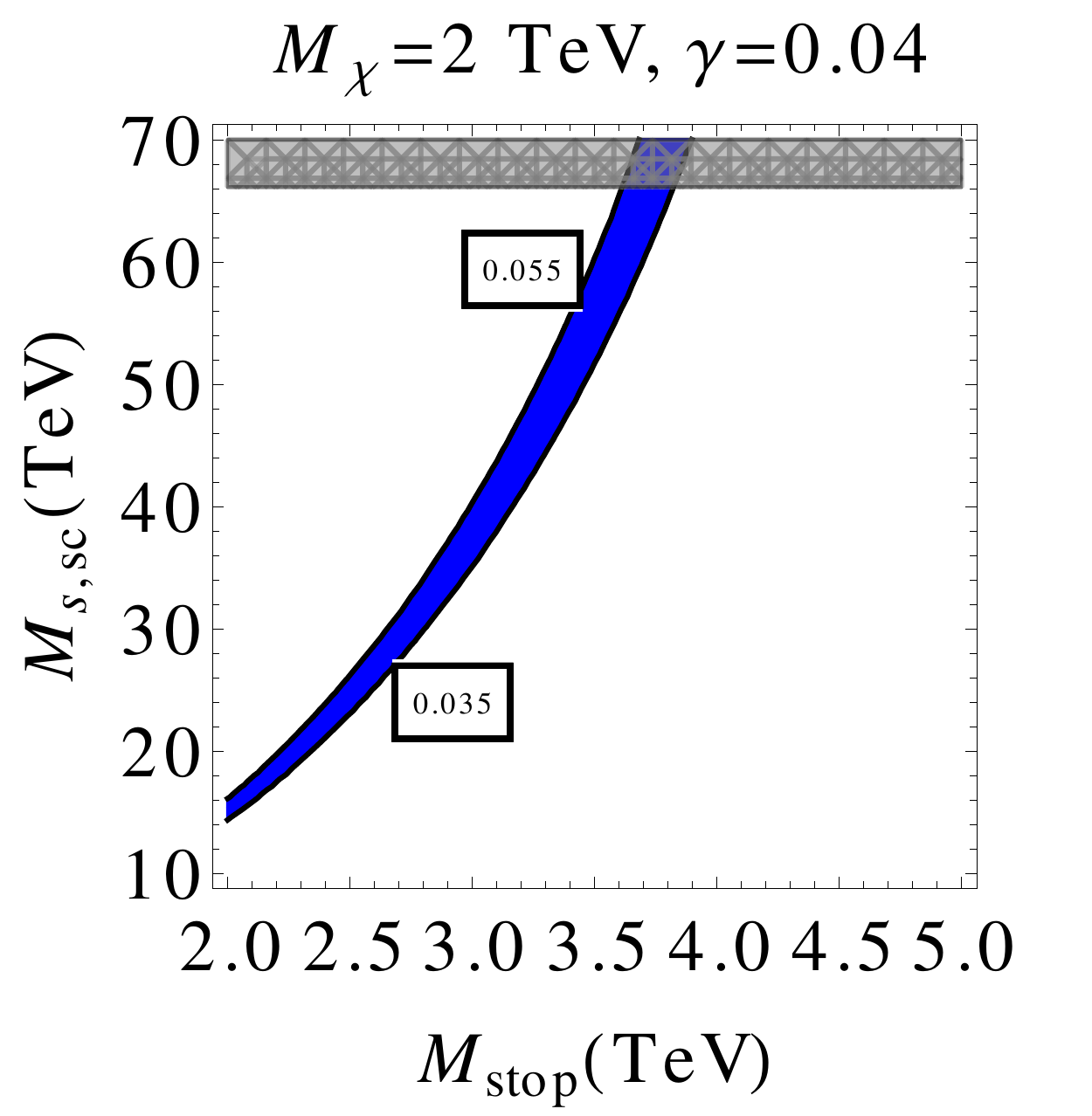}}
\caption{Contour plots of $\Omega_{B}$ as a function of  the coupling $\gamma\equiv\sqrt{\abs{\alpha\beta}}\simeq\sqrt{\abs{\alpha}}$, the mediator mass $M_{\tilde{S}}$ and the stop mass $\tilde{M}$. The blue shaded region corresponds to the observed baryon abundance. The gray shaded region is excluded by the condition of cold freezeout: $M_{\chi}/T_{f}\geq 3$. The plots are obtained by taking $M_{\chi}=2$ TeV, $\mu=1.7$ TeV, $M_{\tilde{B}}=3.5$ TeV, $M_{S}=3$ TeV, and $O(1)$ phases of $\epsilon$ and $\beta$. Furthermore, in a) $M_{\tilde{S}}=30$ TeV is assumed. In b) the coupling constant is fixed at $\gamma=0.04$.}
\label{fig:abundance2}
\end{figure}

The results of the numerical analysis are similar to the ones of the first model and they are presented in Fig.~\ref{fig:abundance2}. In order to reproduce the observed baryon abundance, the stop must have a mass $\tilde{M}\lesssim 4$ TeV, when the baryon parent has a mass of about $2$ TeV. Once again we find that $\chi$ must interact very weakly, $0.02\lesssim\gamma\equiv\sqrt{\abs{\alpha\beta}}\lesssim 0.1$, and that its annihilation must be mediated by a heavy scalar $10~\text{TeV}\lesssim M_{\tilde{S}}\lesssim 60$ TeV.

\begin{figure}[h]
\centering
\includegraphics[scale=0.6]{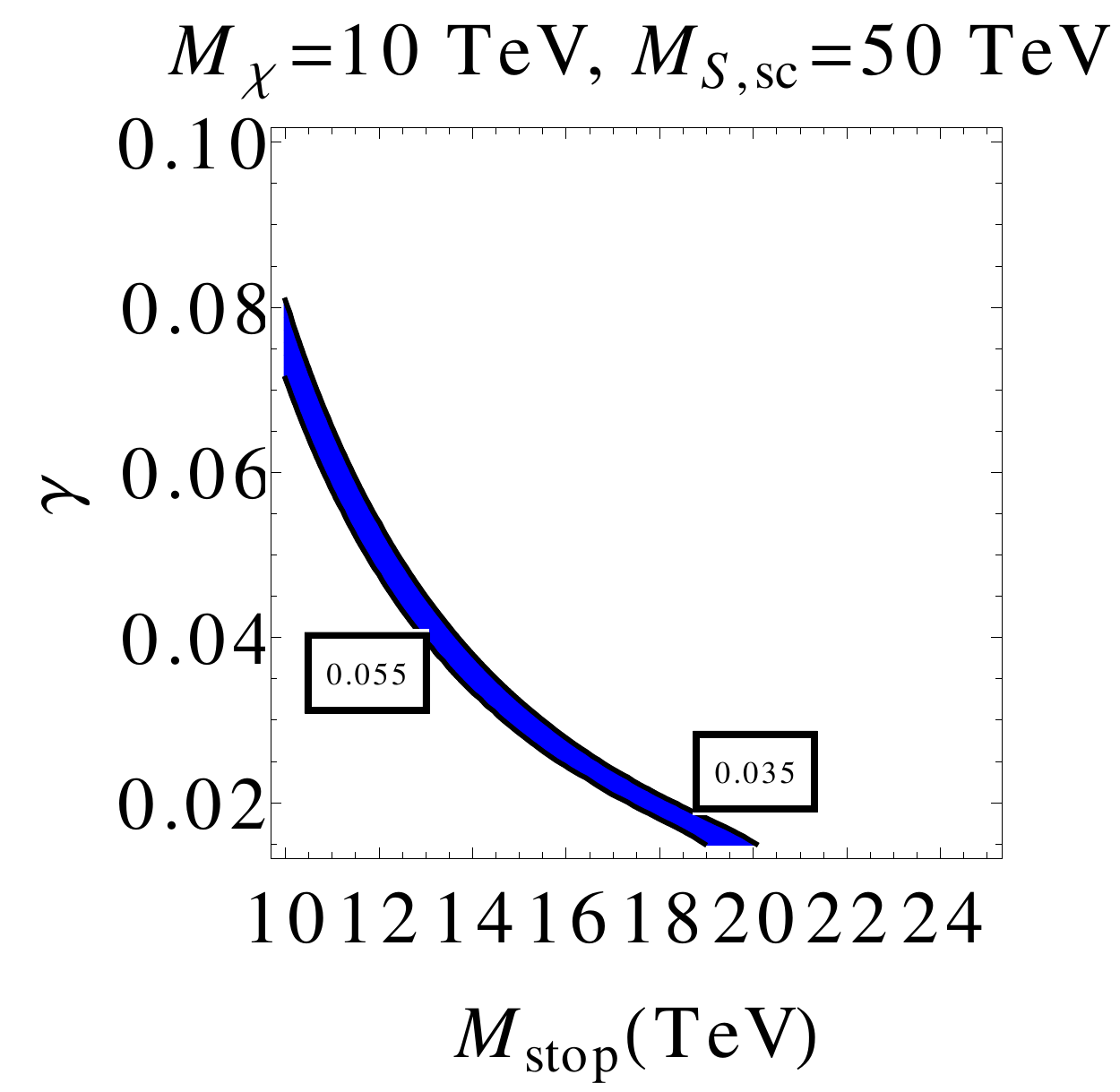}
\caption{Contour plot of $\Omega_{B}$ as a function of the annihilation coupling and the stop mass for heavier baryon parent: $M_{\chi}\simeq 10$ TeV. The blue shaded region corresponds to the observed baryon abundance. The plot is obtained taking $M_{\tilde{S}}=50$ TeV, $\mu=8$ TeV, $M_{S}=15$ TeV, $M_{\tilde{B}}=3.5$ TeV and $O(1)$ phases for $\epsilon$ and $\beta$. In the plotted region of $\gamma$ and $\tilde{M}$ there are no constraints coming from the requirement of cold freezeout.}
\label{fig:abundance210}
\end{figure}

An advantage of this model is that it can more easily accommodate a heavier stop. Indeed, from (\ref{eq:asy2}), it is evident that raising $\mu$ raises also the upper bounds on the stop mass. In order to take a larger $\mu$, also $M_{\chi}$ has to be raised. In Fig.~\ref{fig:abundance210} we show the constrains on the parameter space for $\mu\lesssim M_{\chi}\simeq 10$ TeV, $M_{\tilde{S}}\simeq 50$ TeV, and keeping the other parameters as in the previous analysis. We see that $\tilde{M}\lesssim 20$ TeV, with $\gamma\simeq 10^{-2}$. One may also consider a heavier baryon parent, allowing for heavier stops, without necessarily tuning the coupling $\gamma$. However, as we mentioned in Sec.\ref{sub:first}, in this region of parameter space the connection with the $\Omega_{B}-\Omega_{DM}$ coincidence is weakened.

\section{Conclusions}
\label{sec:conclusions}

The explanation of the BAU  is one of the main open problems in Particle Physics. In this paper we focused on those mechanisms of Baryogenesis which involve the $\slashed{B}$ couplings of the MSSM with R-parity violation. We reviewed the theorem of Nanopoulos and Weinberg, providing a detailed discussion of its assumptions. In particular the result applies to decaying particle which are stable when the $\slashed{B}$ interactions are switched off. We then examined some of the existing scenarios of Baryogenesis and Leptogenesis through RPV, in light of the aforementioned result. We provided examples of models where the theorem seems not to be taken into account.

We then focused on the possibility that the observed baryon abundance might be connected to the would-be abundance of a metastable WIMP, as proposed in \cite{Cui:2012jh}. This scenario can naturally explain the coincidence between $\Omega_{DM}$ and $\Omega_{B}$ if the mass and the annihilation coupling of the metastable particle are similar to those of a cold dark matter candidate.
Starting from the model in \cite{Cui:2012jh}, we investigated two possible realisations in SUSY with R-parity violation, with the field content of the MSSM enriched by only two chiral superfields. One of them contains a metastable Majorana fermion which decays after freezeout into baryons. A heavier squark mediates this decay through RPV couplings. The other one contains a scalar which provides an annihilation channel for the baryon parent. Only the B-violating interactions $\lambda^{''}_{ijk}$ were considered.
We computed the CP asymmetries produced in two different decay channels of the baryon parent. The parameter space in both cases is constrained by the requirement that the obtained baryon abundance reproduces the observed value $\Omega_{B}\approx 0.05$ (Planck). A further assumption which restricts the available parameter space is that the metastable particle undergoes freezeout when non-relativistic. The latter requirement is imposed in the philosophy of explaining the $\Omega_{B}-\Omega_{DM}$ coincidence, as in \cite{Cui:2012jh} where Dark Matter is assumed to be a cold WIMP.

In the first realisation, the baryon parent decays through mixing with a higgsino, then through the R-parity violating couplings to SM states. Therefore the leading decay channel is mediated by an heavy stop. The CP asymmetry is suppressed by loop factors, and by the separation between the stop mass and the electroweak scale. Assuming that the metastable particle has a TeV-scale mass, the stop mass is constrained to be in the multi-TeV region. 

The second decay channel involves another Majorana fermion, the superpartner of the annihilation mediator. Once again a heavy stop is involved in the decay to SM states, through R-parity violating interactions. 
As a consequence of mixing between binos and higgsinos, the CP asymmetry is suppressed by the separation between the stop mass and the $\mu$ mass of the MSSM. The latter must be smaller than the mass of the metastable particle, otherwise the decay channel is forbidden. Considering a TeV-scale mass for the baryon parent, the stop mass can again be at most at the multi-TeV scale.

In order to reproduce the observed baryon abundance, the parameters which determine the annihilation cross section of the metastable particle are rather constrained. In particular, the characteristic coupling $\gamma\equiv\sqrt{\abs{\alpha\beta}}$ is required to be very weak $\gamma\sim 10^{-2}$, and the mediator must be rather heavy $M_{\tilde{S}}\sim 10$ TeV. Both models can accommodate a heavier stop. This however requires to raise the mass of the baryon parent and $\mu$. In particular, in both cases we found $\tilde{M}\lesssim 20$ TeV for $M_{\chi}\simeq 10$ TeV and $\mu$ in the multi-TeV region, with annihilation parameters in the range mentioned above. We would like to remark that stop masses far from the TeV scale can still lead to the observed baryon abundance, though they require values of the other parameters which are less well-motivated from the point of view of the $\Omega_{DM}-\Omega_{B}$ coincidence. 

The implications for the explanation of the observed coincidence between dark and baryonic matter, $\Omega_{DM}\simeq 5 \Omega_{B}$, are as follows. In the models that we have examined, the metastable parent is required to annihilate with a coupling which is at least one order of magnitude smaller than the weak coupling. The latter is known to provide the observed Dark Matter abundance, when associated to the annihilation of a TeV-scale cold relic. Therefore these models require a moderate tuning of parameters in order to explain the coincidence between $\Omega_{B}$ and $\Omega_{DM}$ in the framework of WIMP DM.

Let us finally mention an important implication concerning displaced vertices at LHC (see \cite{Graham:2012th, Barry:2013nva} for recent discussions of displaced vertices from SUSY, \cite{Aad:2011zb} for experimental searches). In the framework that we considered, the metastable particle is very long lived. Indeed its decay length is determined by the B-preserving channel $\chi\rightarrow H,\tilde{H}$. As a consequence of the smallness of the coupling $\epsilon$, we find $l_{D}>1$ cm, for a TeV-scale baryon parent. This implies that such a particle, if produced, would leave a displaced vertex inside the detector.   

\subsection*{Aknowledgements}

I thank R. Rattazzi for enlightening and helpful discussions. I am indebted to P. Lodone for invaluable help. I also thank F. Riva and M. Nardecchia for useful comments, and L. Witkowski and A. Hebecker for helpful suggestions on the structure of this paper. This work was partially supported by DFG Graduiertenkolleg GRK 1940 "Particle Physics Beyond the Standard Model".

\vspace{0.3cm}



\end{document}